%% file: article.tex
\documentclass[aps,prb,reprint,amsmath,amssymb,superscriptaddress,floatfix]{revtex4-2}

\usepackage{xcolor}
\usepackage{graphicx}
\usepackage{dcolumn}
\usepackage{bm}
\usepackage[colorlinks, citecolor=blue, linkcolor=blue]{hyperref}

\usepackage{graphicx}
\usepackage{dcolumn}
\usepackage{bm}
\usepackage{hyperref}

\usepackage{xcolor,colortbl}
\usepackage{tikz,xcolor,hyperref}
\usepackage{orcidlink}
\definecolor{lime}{HTML}{A6CE39}
\usepackage{diagbox}
\usepackage{makecell}
\usepackage{nicematrix}

\usepackage{orcidlink}
\usepackage{txfonts}

\usepackage[]{graphicx}      
\usepackage{bm}             	
\usepackage{times}			
\usepackage{tabularx}
\usepackage{color}
\usepackage{dcolumn}		
\usepackage{amsmath}
\usepackage{amssymb}
\usepackage{amsfonts}
\usepackage{times}
\usepackage{tabularx}
\usepackage{xcolor,colortbl}
\usepackage{amsmath}
\usepackage{tikz,xcolor,hyperref}
\usepackage{blindtext}
\usepackage{hyperref} 
\UseRawInputEncoding

\begin{document}
 
\title{Method of analysis of the spectra obtained by microfocused Brillouin light scattering}

\author{N. Benaziz \orcidlink{0009-0006-2054-1957}}
 \affiliation{%
Universit\'e Paris-Saclay, CNRS, Centre de Nanosciences et de Nanotechnologies, Palaiseau, France
}%

\author{T. Devolder \orcidlink{0000-0001-7998-0993}}
\email{thibaut.devolder@cnrs.fr}
\affiliation{%
Universit\'e Paris-Saclay, CNRS, Centre de Nanosciences et de Nanotechnologies, Palaiseau, France
}%
\author{J-P. Adam  \orcidlink{0000-0003-2025-7105}}
 \affiliation{%
Universit\'e Paris-Saclay, CNRS, Centre de Nanosciences et de Nanotechnologies, Palaiseau, France
}%

\date{\today}

\begin{abstract}
Brillouin Light Scattering is a powerful technique to measure the microwave excitations present in a magnetic system. In microfocused mode, the light is focused on the sample using a microscope objective. This accelerates substantially the measurement but results in mixing the response of all microwave excitations into a single spectrum, such that modeling is required to disentangle the contribution of each spin wave. By assuming that a spectrum collected in microfocused mode can be approximated by the sum of all back-scattering spectra that can be collected by the microscope objective, we develop a simple and direct way of interpreting spectra. The model is compared to experimental data collected on a 50 nm thick CoFeB magnetic film. The model allows the understanding of the influence of the optical properties of a sample, of the dispersion relation of the spin wave eigenexcitations and of their thickness profiles, as well as of their populations onto the magnitudes and the lineshapes of the characteristic features of a spectrum. 
\end{abstract}

\maketitle

\section{\label{sec:level1}Introduction\\}
Magnons are the resonant low energy disturbances of the magnetic order parameter. Brillouin Light Scattering (BLS, \cite{hillebrands_progress_1999}) is a  well established technique for their study \cite{nembach_effects_2011, roussigne_brillouin_1995, belmeguenai_interfacial_2015}. BLS is typically implemented in wavevector-resolved mode (k-BLS). There, one analyzes the frequency shift of the photons of a parallel beam that are inelastically back-scattered by the magnons of a magnetic sample. For each optical incidence, a k-BLS spectrum can be straightforwardly used to deduce the spin wave (SW) frequencies for the wavevector defined by the measurement geometry \cite{sebastian_micro-focused_2015, gaier_brillouin_2009,hamrle_determination_2009,tacchi_strongly_2019}. The SW dispersion relation can then be constructed by varying the wavevector of the magnon scatterers simply by scanning the sample orientation with respect to the light beam \cite{belmeguenai_interfacial_2015}. The drawback of BLS is that the low scattering cross section of photons by magnons renders this method slow and cumbersome.

One can circumvent this difficulty by instead focusing the light beam on the magnetic sample. This aggregates, within a single experimental configuration, all the orientations of the light beam with respect to the sample, thereby collecting the scattering events from all magnons simultaneously. This so-called "microfocused" BLS spectroscopy ($\mu$-BLS , \cite{demidov_radiation_2004,sebastian_micro-focused_2015}) allows a much faster characterization; unfortunately a direct interpretation of the scattering spectra is difficult, since the spectra result from the interplay between the spin wave properties, the magneto-optical properties and optical diffraction. The past descriptions of $\mu$-BLS spectroscopy \cite{che_multipole_2024}  
do not apply in the general case \cite{freeman_brillouin_2020}. A noticeable exception is the recent work of Wojewoda et al. \cite{wojewoda_modeling_2024-1} who attempted to quantitatively calculate $\mu$-BLS spectra by including all magnetic and optical relevant phenomena; their study relies on a mathematical formalism that implies several steps of numerical integration. This hinders the physical understanding so that it is difficult to anticipate the influence of the SW density of states, of the thickness profiles of the spin waves, and of its interplay with the optical properties of the sample. 
Since the microfocused BLS spectroscopy is getting ever more popular, it is of utmost importance to describe it using an explicit and transparent formalism that takes into account the most physically relevant phenomena.

Our present goal is to build upon the published works of k-BLS \cite{buchmeier_intensity_2007,hamrle_analytical_2010,zivieri_stokesanti-stokes_2002} to formulate a simple but comprehensive model of microfocused BLS spectra, taking into account the full complexity of the magnetic response, and the optical properties of the sample. The key assumption in our analysis is that scattering of photons in directions other than back-scattering can be disregarded. We consider that a spectrum collected in microfocused mode is at first order the sum of all back-scattering events that lie inside the numerical aperture of the microscope objective.
We test our model on a 50 nm-thick $\mathrm{Co}_{40}\mathrm{Fe}_{40}\mathrm{B}_{20}$ film that comprises four families of spin waves contributing to the $\mu$-BLS signal in the 50 GHz-wide frequency range. Our formalism requires the prior knowledge of the magnetic eigenmodes of the system, which can be obtained using experimental data, analytical expressions or numerical eigensolvers.  Our model clarifies the relative magnitudes of the signals arising from the different SW families, as well as the shape of their spectral signatures. It also explains the role of the optical and magneto-optical indices and their impact on the Stokes/antiStokes asymmetry. Guidelines to interpret microfocused BLS spectra can thus be drawn.

The paper is organized as follows. Section \ref{exp} reports an experimental spectrum of a 50 nm thick CoFeB film measured by $\mu$-BLS. It identifies the main spectral features that we will try and understand. Section \ref{sec:presentationmodel} describes step by step the calculation of microfocused BLS spectra from the spin wave properties, the sample optical properties and the geometry of the experimental setup. Section \ref{sec:CoFeBapp} implements the model for a 50 nm-thick CoFeB film. For didactic purposes, this is first done on a subset of the spin waves of the system. This evidences the links between the $\mu$-BLS spectrum and the SW density of states. We then discuss the role of the optical properties of the magnetic material, before calculating the total response from the full set of SWs present in the system and discussing the agreement with experimental data.

\section{Microfocused Brillouin Light Scattering experiments} \label{exp}
\begin{figure}     \centering   
    \includegraphics[width =6cm]{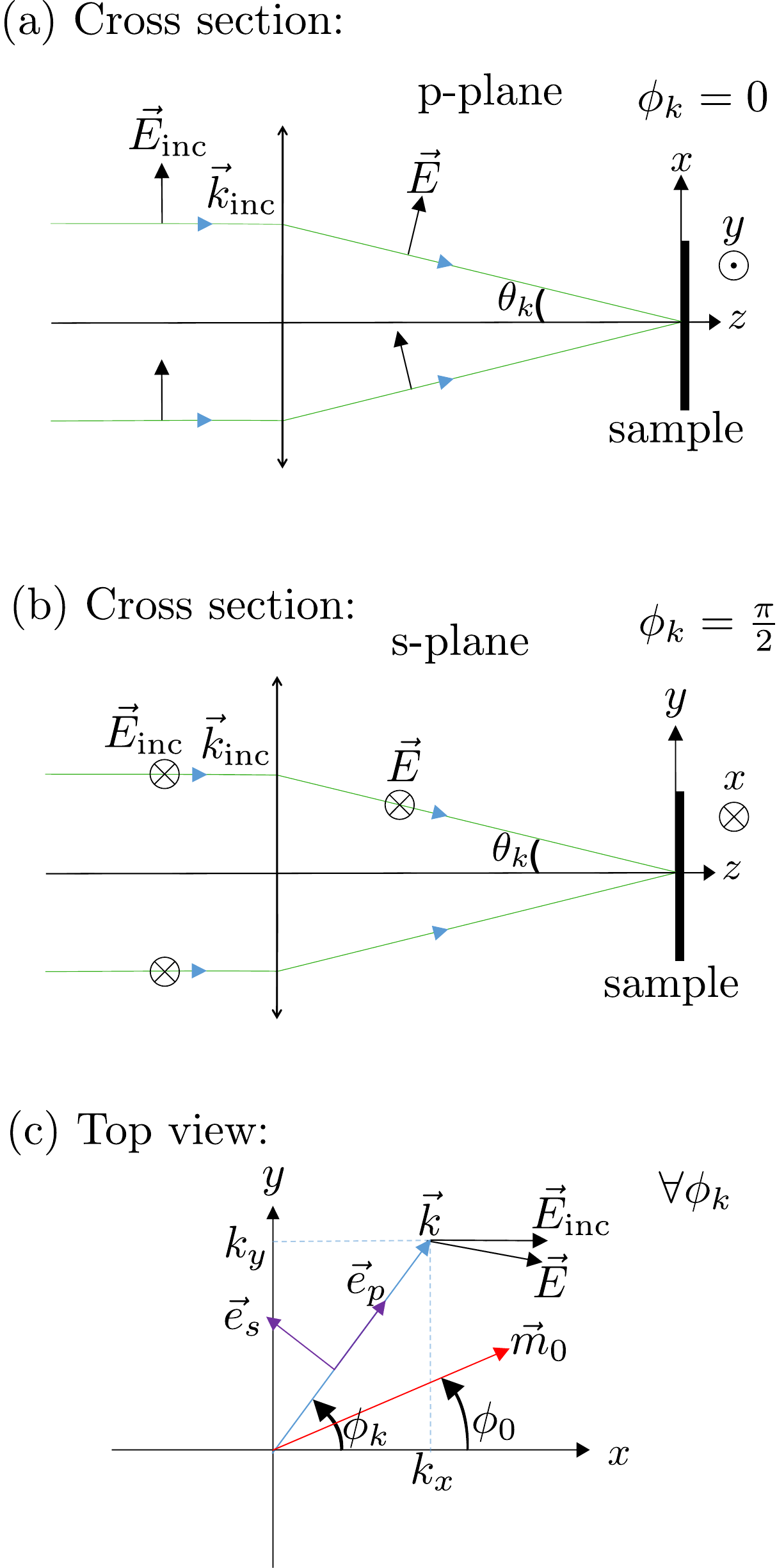}
    \caption{Geometry and reference frames in cross-section view  (a) for $\phi_k$=0 and (b)  for $\phi_k$=$\frac{\pi}{2}$ and (c) top view for $\forall \phi_k$. A parallel light beam with linear polarisation $\vec E_\textrm{inc}$ along (x) is focused by an objective lens on the magnetic sample, with incidence angle $\theta_k$ and rotated polarization $\vec E$. $\phi_k$ is the angle between the incoming polarisation and the wavevector of the spin waves involved in antiStokes processes. $\vec e_p$ is the vector of the incident plane perpendicular to the light wavevector, and $\vec e_s$ is the vector perpendicular to the incident plane. $\phi_0$ is the angle between the equilibrium magnetization $\vec m_0$ and the $x$ axis.}
    \label{shcmconvt}
\end{figure}
\begin{figure*}
    \centering
    \includegraphics[scale=0.1,width =17cm]{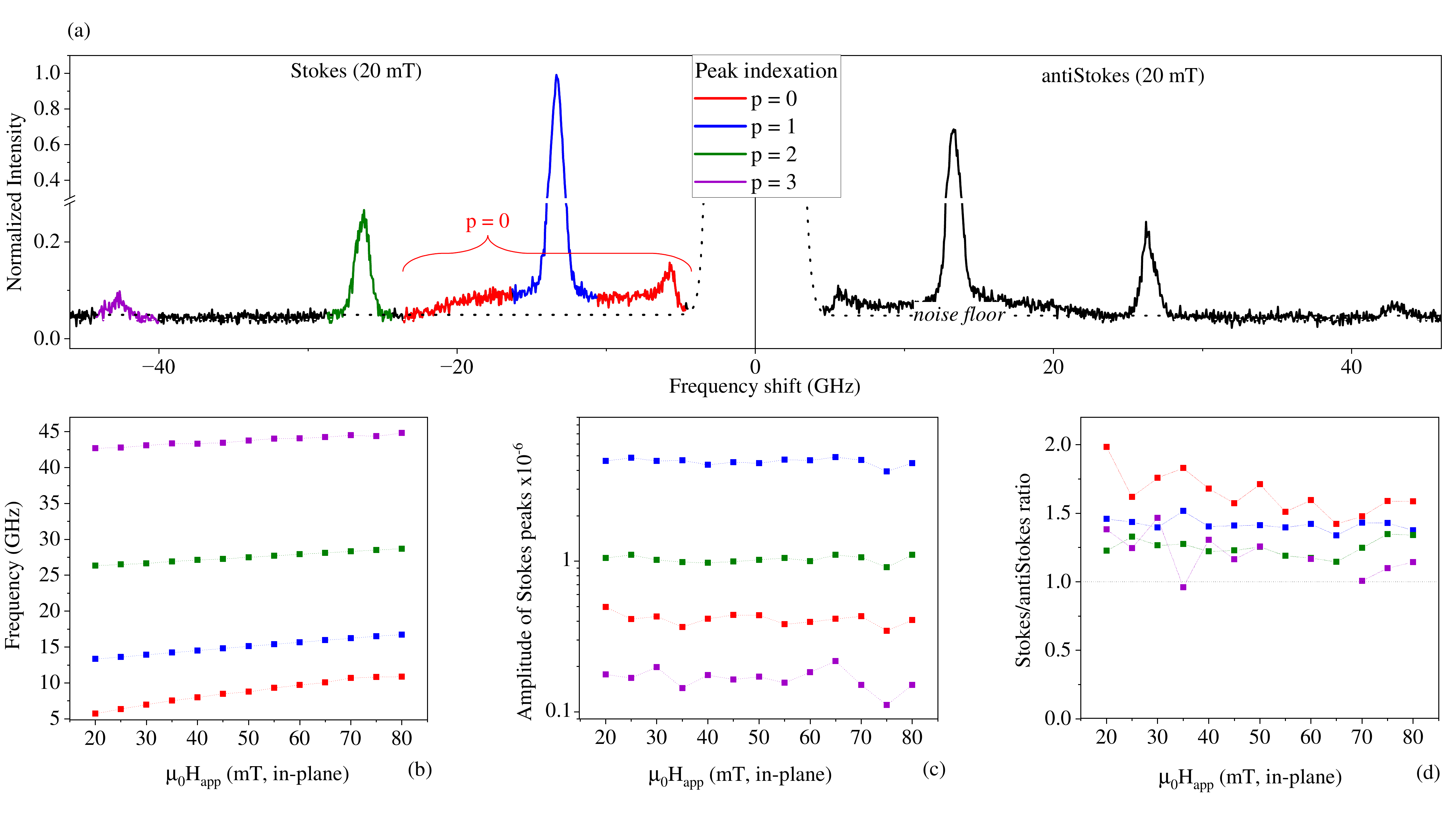}
    \caption{Experimental results: (a) microfocused BLS spectrum recorded for an applied field of $\mu_0 H_\textrm{app} = 20$ mT. The peaks are normalized with respect to the Stokes p1 peak.  (b) Field dependence of the frequencies of the maxima of the spectra (c) Field dependence of the noise-floor-corrected intensity maxima of the Stokes peaks. Intensities are shown on a logarithmic scale. The intensities are normalized to the elastic peak  (d) Field dependence of the ratios of the noise-floor-corrected maxima of the intensity of the Stokes versus antiStokes parts of the spectra for the 4 peaks. In (b),(c) and (d), the dashed lines are guides to the eye. 
}
    \label{resultscofeb}

\end{figure*}

The microfocused Brillouin Light Scattering experiments were performed using a high numerical aperture NA = 0.75 microscope objective to focus an incoming linearly polarized ($\vec E_\textrm{inc} \parallel \vec e_x$, Fig. \ref{shcmconvt})  green laser beam (wavelength $\lambda$ = 532 nm, 2.34 eV) onto the sample surface. After interacting with the sample, the outgoing photons pass through an analyzer that selects those with a polarization along $\vec e_y$. The frequency shift of the outgoing photons is measured using a three-pass Fabry-Perot interferometer \cite{mock_construction_1987}. The sample is a Ta/$\mathrm{Co}_{40}\mathrm{Fe}_{40}\mathrm{B}_{20}$ (d = 50 nm) / Ta (3 nm, cap) magnetic film grown by sputter deposition on a $\mathrm{SiO}_2$/Si substrate.  The applied field is parallel to the incident laser polarization.

A representative subset of our experimental results is reported in Fig.~\ref{resultscofeb}. For applied fields $\mu_0 H_\textrm{app} < 80$ mT, the spectra [Fig.~\ref{resultscofeb}(a)] systematically comprise four peaks that emerge from the noise floor in the frequency window $|f| \in [2, 45]$ GHz. For reasons that will be clarified in section III, we shall index these peaks as $p=0,...,3$ and refer to them as arising from the branch of spin waves that have the most thickness-uniform profiles [FerroMagnetic Resonance (FMR), Magnetostatic Surface Spin Waves (MSSW) and Backward Volume Spin Waves (BVSW)], or arising from the branches of spin waves that have a Perpendicular Standing Spin Wave (PSSW) character ($p>0$).
The field-dependence of the frequencies of our four peaks bears a large similarity to that typically measured in ferromagnetic resonance experiments: the frequency of the $p=0$ peak looks alike that of a Kittel mode and the frequencies of the next peaks are similar to PSSW modes [see Fig.~\ref{resultscofeb}(b)].
The amplitude of the $\mu$-BLS signal is not monotonic with the peak index: the second peak  $(p=1)$ is by far more intense than all others [see Fig.~\ref{resultscofeb}(c)] and is constant with the applied field. Besides, the signal detected on the Stokes side of the spectrum (frequency shift $f < 0$, corresponding to magnon creation) is systematically larger than that detected on the antiStokes side ($f > 0$, corresponding to magnon annihilation). The Stokes/antiStokes (S/AS) asymmetry, defined as the ratio of the maxima of the signals at $f < 0$ divided by the maxima at $f > 0$, is plotted in Fig.~\ref{resultscofeb}(d). It is typically greater than unity. 
There are several striking features to be noticed in Fig.~\ref{resultscofeb}.  \begin{itemize}
\item The evolution of the amplitude of the $\mu$-BLS signal with the successive peaks is quite unexpected. It is in stark contrast with the evolution typically obtained by inductive measurement techniques such as vector network analyzer ferromagnetic resonance (FMR), where the signal of the FMR mode has an intensity that exceeds far above that of the next ones  \cite{bilzer_study_2006}.
\item The shape of the $p=0$ peak is substantially skewed to the high frequency side. It seems to emerge from a shallow hill-like broad peak. In contrast with the usual case of FMR, in which symmetric lines are most often encountered.
\item The $p=1-3$ peaks have a much more symmetrical appearance. However, they do not have a perfect Lorentzian lineshape. Their linewidths (FWHM) are 0.8 GHz for $p$ = 1, 1.0 GHz for $p$ = 2 and $1.3\pm0.3 $ GHz for $p$ = 3, which are much larger than the Gilbert linewidths observed in FMR on the same samples (not shown, typically 0.2 GHz for $p=1$).
\end{itemize}
The next section details the model used to quantitatively construct microfocused BLS spectra.

\section{\label{sec:presentationmodel} Scattering spectrum of a population of spin waves}
\subsection{\label{{subsec2:geometryandnotation}} Geometry and notations}

The notations are summarized in Table~\ref{tab1}. We express the magnetization as: 
\begin{equation*}
    \vec{m}(\vec r, t)= \vec m_0(\vec r) + \vec{\delta  m} (\vec r,t),
\end{equation*}
with $\vec m_0(\vec r)$ being the normalized direction of the equilibrium magnetization supposed in the $xy$ plane and $\vec{\delta m}(\vec r,t) \perp \vec m_\mathrm{0}$ being the dynamic magnetization. The index "0" will be used whenever referring to an equilibrium quantity. In the linear limit, the dynamic magnetization can be written in frequency-domain by defining: 

\begin{equation*}
    \vec{\delta  m}(\vec  r,t)=\sum_{\Psi_{k}}  \mathrm{Re}\Big(\tilde b_k(t) \vec{\tilde{m}}_{\Psi_k} (\vec r)e^{-i\omega_{\Psi_{k}} t} \Big),
\end{equation*}
with $\vec{\tilde{m}}_{\Psi_k} (\vec r)$ being the complex \footnote{The tilde recalls that they are complex quantities. } spatial mode profiles of the spin wave (SW) modes $\Psi_{k}$ \cite{massouras_mode-resolved_2024,daquino_novel_2009}. The  $\tilde b_k (t)$  are the spin wave complex amplitudes (see annex I). Since there are several SWs with the same wavevector $\vec k$, we name the SW eigenmode $\Psi_{k}$ instead of referring to it with just the label $\vec k$.  Each $\vec{\tilde{m}}_{\Psi_k}(\vec r)$ is normalized to correspond to the dynamic magnetization of a \textit{single} magnon per a given unit volume. The normalization procedure is described in Annex I. 
The spin waves $\Psi_{k}$ have wavevectors $\vec k=\{k_x, k_y\}$, assumed to lie in the sample plane $xy$. Their frequencies are $\omega_{\Psi k}$ and their linewidths are $\Delta \omega_{\Psi k}$, written for short $\omega_{k}$ and $\Delta \omega_k$.
The spatial mode profiles of each SW mode can be written in the spherical coordinate system $(\vec e_r (\vec r) \parallel \vec m_0 (\vec r), \,\vec e_{z}(\vec r), \,\vec e_{\phi_0}(\vec r))$ attached by the local equilibrium magnetization $\vec m_0 (\vec r)$:   
\begin{equation}
\vec{\tilde{m}}_{\Psi_k}(\vec r)= {e^{i \vec k .\vec r}} \Big(\tilde m_{\Psi_k,{\mathrm{{\phi_0}}}} ~\vec e{_\mathrm{{\phi_0}}}(\vec r) +\tilde m_{\Psi_k,{\mathrm{{z}}}} ~\vec e{_\mathrm{{z}}}(\vec r) \Big).
\label{mphil}
\end{equation}

We need to introduce additional notations to describe how the SW interacts with the focused laser beam of a $\mu$-BLS setup.
The figure \ref{shcmconvt} presents the geometry. A Gaussian laser beam of  wavevector $\vec{k}_{\mathrm{inc}}$ propagates along the $z$ axis with linear polarization $\vec{E}_{\mathrm{inc}}$ along $x$.  A high numerical aperture lens focuses the light beam into a cone with the focal point at the sample surface at $x=y=z=0$. The numerical aperture of the microscope defines the maximal conical angle of incidence $\theta_\textrm{max}$: $$\theta_\textrm{max} =\arcsin(\text{NA})\approx 49^\circ.$$
Since we consider the back-scattering of photons, each ray inside the incident cone can be indexed by two angles $\{\theta_k, \phi_k \}$ and shall interact through an antiStokes process with the SWs possessing the wavevector such that: 
\begin{equation}
    \left\{
    \begin{array}{ll}
        \theta_k=\theta(k_x, k_y) = \arcsin{\left(\frac{\lambda}{4\pi}\sqrt{k_x^2 + k_y^2}\right)} \\
        \phi_k = \phi(k_x, k_y) = \arg(k_x + i k_y)
    \end{array}
    \label{angle}
    \right. 
\end{equation}

We emphasize that $\theta_k$ and $\phi_k$ describe the spin wave wavevector at play for the back-scattering of a photon. $\theta_k$ and $\phi_k$ are \textit{not} the direction of the light beam that interacts with the spin wave. The orientation of the spin wave wavevector $\phi_k$ matches with the direction of the light beam \textit{only} for antiStokes processes (photon scattering with magnon annihilation and positive frequency shift of the light), where the prior existence of a magnon is associated with a quasistatic diffraction grating at the surface of the material. 

The wavevector range of probed magnons is set by $\theta_{\mathrm{max}}$ and is bounded by: $$k_{\mathrm{max}}=\frac{4\pi}{\lambda} \mathrm{NA} \approx~18~ \mathrm{rad}.\mathrm{\mu m^{-1}}.$$ \input{table_input_model}

For the Stokes process (photon scattering with magnon creation and negative frequency shift of the light) involving the very same spin wave with its $ \phi_k$ and $\theta_k$, the involved light beam has the same incidence angle $\theta_k$ but is oriented in the direction \textit{symmetric} with respect to $z$, i.e. it is oriented at $\phi_k+\pi$.
For a spin wave $\Psi_k$ with its $\theta_k$ and $\phi_k$, the Stokes processes can be modeled by making the time-reversal transformation \cite{zivieri_stokesanti-stokes_2002}: 
\begin{equation}
    \left\{
    \begin{array}{lll}
        \textrm{antiStokes} & \rightarrow &\textrm{Stokes} \\
        \omega >0 &  \rightarrow & - \omega  \\
         \vec{\tilde{m}}_{\Psi_k}(\vec r) &\rightarrow &\vec{\tilde{m}}_{\Psi_k}^*(\vec r)
    \end{array}
    \label{AStoStransformation}
    \right. 
\end{equation}
where the star means complex conjugation.

In addition to bending the light rays, the focusing reorients their polarization. 
From Eq.~3.51 of Ref.~\cite{novotny_principles_2012}, each light ray inside the cone is described as a sum of p-polarized and s-polarized light [see Fig. \ref{shcmconvt}(b)], with the electric field being,
\begin{equation}
\vec {E}(\vec k)=\sqrt{\cos\theta_k} \, E_{\mathrm{inc}}(\vec k)\Big({E_p(\vec k)}\vec e_p+{E_s(\vec k)}\vec e_s\Big).
\label{vecE}
\end{equation}
The unit vectors of the local reference frame are:
\begin{equation}
    \left\{
    \begin{array}{ll}
        \vec{e}_p=\cos\phi_k \cos\theta_k ~\vec e_x +\cos\theta_k \sin\phi_k ~\vec e_y-\sin\theta_k ~\vec e_z\\
         \vec{e}_s=-\sin\phi_k~\vec e_x+\cos\phi_k~\vec e_y
    \end{array}
    \right.
    \label{epesvect}
\end{equation}
and the two components of the light polarization are \footnote{The conservation of the energy of the light beam upon Eqs.~\ref{vecE}-\ref{EpEs} can be seen from the fact that the tilted beam irradiates a sample surface enlarged by the factor $\cos\theta_k$ and from the statement that $\forall \phi_k$, we have $||\vec {E}(\vec k)||= {\cos\theta_k} \, ||\vec E_{\mathrm{inc}}||$.
}:
\begin{align}
    \centering
        E_p=\cos\phi_k \quad\text{and}\quad  E_s=-\sin\phi_k
    \label{EpEs}
\end{align}

\subsection{\label{sec:setup} Description of the model }
In this model, we will consider that the skew-scattering of photons (i.e. scattering in directions other than back-scattering) can be disregarded. Under this assumption, the $\mu$-BLS spectrum $F_\textrm{bls}(\omega)$ collected by the lens is the sum of the back-scattering contributions of all SW modes $\Psi_k$ present in the system and that involve light rays that are inside the numerical aperture of the microscope objective: 
\begin{equation}
    F_{\mathrm{bls}}(\omega)=\sum_{\Psi_{k}} \eta(\Psi_k)\, R(\vec k)\, I(\Psi_k) \,\mathcal L_{\Psi_k} (\omega) \label{BLSspectrumMasterEquation}
\end{equation}
where \begin{itemize}
    \item 
$\eta(\Psi_k)$ is the population of the $\Psi_k$ magnon,  \item $R(\vec k)$ is an instrumental function related to the light diffraction (Rayleigh criterion), \item $I(\Psi_k)$ is the intensity scattered by each magnon, \item
$\mathcal L_{\Psi_k}(\omega)$ accounts for the finite lifetime of the magnon modes, and is expressed by the Lorentzian function: 
\begin{equation}
\mathcal L_{\Psi_k} (\omega)=\frac{\Delta \omega_{\Psi_k}}{2\pi} \frac{1}{\left({\Delta \omega_{\Psi_k}}/{2}\right)^2 + (\omega - \omega_{\Psi_k})^2} \label{lorentz} \end{equation}
which has a unit integral. 
\item If the spectrometer has a finite spectral resolution $\xi \neq 0$, the linewidth should be taken as an \textit{effective} linewidth \cite{cochran_calculation_1988} being: $\Delta\omega_{\Psi_k}'=\sqrt{\Delta\omega_{\Psi_k}^2+\xi^2}$.

\end{itemize}

Depending on the system under investigation, the population $\eta(\Psi_k)$ can either be a thermal population, or any out-of-equilibrium population, for instance as driven by a forcing stimulus through the mode susceptibility. We will focus on the thermal equilibrium case when $\eta(\Psi_k)$ is simply proportional to the Bose-Einstein distribution \footnote{Because of Gilbert damping and of multimagnon processes like three-magnon scattering, the number of magnons is not a conserved quantity, even at thermal equilibrium. As a result, its conjugate variable --the chemical potential-- can be taken as zero and the Bose-Einstein factor can be taken as equal to the Planck distribution.}:
$    \eta({\Psi_k}) \propto [{\exp\frac{\hbar\omega_{\Psi_k}}{k_B T}-1]^{-1}}$
where $k_B$ is the Boltzmann constant 
and $T$ is the temperature. 


From Eq.~3.9 of Ref.~\cite{novotny_principles_2012} the instrumental function accounting for the diffraction limit ("Rayleigh criterion") is \footnote{The above Eq.~\ref{instrument} is an approximation valid only in the paraxial regime of geometrical optics. For strongly focused polarized light, higher order effect leads to a focal spot that is slightly elongated in the direction of polarization, such that this function should be written as $R(k, \cos 2\phi)$ with a more complicated expression (see Eq. 3.58, 3.60 and 3.66 of Ref. \cite{novotny_principles_2012}). The approximation done in Eq.~\ref{instrument} has little consequence in practical situations, since as we shall see later in Eqs.~\ref{phiS} and \ref{phiP}, the sensitivity drops dramatically at the large k's where a slight modification Eq.~\ref{instrument} would start to make sense.}: 
\begin{equation}
    R(k)\approx\exp\Big(-8{\frac{|\vec k|^2}{k_{\mathrm{max}}^2}}\Big)
    \label{instrument}
\end{equation}

The main difficulty within Eq.~\ref{BLSspectrumMasterEquation} resides in the evaluation of the back-scattering contribution $I(\Psi_k)$ of each individual magnon mode $\Psi_k$. 

\subsection{antiStokes back-scattering by individual magnons}
The k-BLS signal $I(\Psi_k)$ is constructed from the light scattered from every slice $dz$ of material along the light propagation path.
For the following, we will consider metallic films of thicknesses $d$ larger than the optical skin depth (typically 20 nm \cite{camley_theory_1981}), or dielectric magnetic films deposited on an antireflection substrate: this allows to assume that all photons to be scattered propagate in the $z>0$ direction and multiple reflections can be ignored. 
As often in magneto-optics, the material's response differs when the light is p or s-polarized. From Refs. \cite{hamrle_analytical_2010, buchmeier_spin_2003}, we can reformulate the electric field of the light scattered for a p-polarized and s-polarized incoming ray as:
\begin{equation}
    \begin{array}{c}
    \vec{  \tilde {E}}^{\mathrm{BLS}}_p{(\Psi_k)} = 
      - E_{p} r_{pp}^{\theta_k} \int_0^d 
           \tilde \Phi_p(z) \, e^{\frac{i 4 \pi \tilde n(\theta_k)  z}{\lambda}}  \, \mathrm{d}z~\vec e_s \\
           \\
    \vec{  \tilde {E}}^{\mathrm{BLS}}_s{(\Psi_k)} = 
       E_{s} r_{ss}^{\theta_k} \int_0^d 
           \tilde \Phi_s(z) \, e^{\frac{i 4 \pi \tilde n(\theta_k)  z}{\lambda}}  \, \mathrm{d}z ~\vec e_p
           \end{array}
    \label{electricfieldps}
\end{equation}
where $\tilde n(\theta_k) =\sqrt{\tilde\epsilon-\sin^2\theta_k}$ is the normalized light wavevector along the $z$ axis \footnote{The square root convention is chosen so that $\mathrm{Im}(\tilde n(\theta_k))>0$. This ensures the decrease with $z$ of the factor $ e^{({i 4 \pi \tilde n(\theta_k) z})/{\lambda}}$, expressing the absorption and phase shift of the light along its penetration path. } and  $ \tilde \epsilon$ is the diagonal term of the relative permittivity of the material. The $4\pi z$ recalls that the light travels back and forth through the depth $z$. The factors $r_{pp}^{\theta_k}$ and $r_{ss}^{\theta_k}$ are the reflection coefficients (the so-called Fresnel coefficients) of p and s-polarized light rays of incidence $\theta_k$. In most materials, the Fresnel coefficients are almost independent from the magnetics. For a semi-infinite material of diagonal and isotropic $\tilde \epsilon$, they would be \cite{mansuripur_physical_1995}:
\begin{equation*}
    \tilde r_{pp}^{\theta_k}=\frac{\tilde n(\theta_k)-\tilde \epsilon\cos\theta_k}{\tilde n(\theta_k)+\tilde \epsilon\cos\theta_k},~ 
   \tilde  r_{ss}^{\theta_k}=\frac{\cos\theta_k-\tilde n(\theta_k)}{\cos\theta_k+\tilde n(\theta_k)}
\end{equation*}
In Eq.~\ref{electricfieldps}, $\tilde \Phi_s(z)$ and $\tilde \Phi_p(z)$ express how the polarizations of the scattered light from initially s or p-polarized incoming rays are modified (added "Kerr" rotation and added "Kerr" ellipticity, in addition to the positive frequency shift) compared to the incoming light, per unit material thickness $dz$ at the depth $z$.
When passing again through the microscope objective, the scattered rays originating from the s-polarised part and the p-polarized part of the incoming beam (Eq.~\ref{vecE} and \ref{EpEs}) recombine  
\begin{equation}
    \vec E^\textrm{BLS}(\Psi_k)=  \vec {\tilde E}^{\mathrm{BLS}}_p(\Psi_k)  +\vec {\tilde E}^{\mathrm{BLS}}_s(\Psi_k)  \label{EdePhSIk}
\end{equation}

After passing back through the lens, the ray encounters the polarizing beam splitter for a second time, where only the light polarization along the $\vec e_y$ axis is kept. Finally, the collected $\mu$-BLS signal to be injected in Eq.~\ref{BLSspectrumMasterEquation} is :

\begin{equation}
    I(\Psi_k)=\Big\| \tilde E^{\mathrm{BLS}}_p(\Psi_k)\cos\phi_k   + \tilde E^{\mathrm{BLS}}_s(\Psi_k)\sin\phi_k \Big\|^2 \label{IdePhSIk}
\end{equation}

\subsection{Relation with magneto-optical Kerr effect} \label{BLSversusMOKE}
The quantities $\tilde \Phi_p$ and $\tilde \Phi_s$ within Eq.~\ref{electricfieldps} are closely related \cite{cochran_calculation_1988,martin_valderrama_layer-resolved_2024,valderrama_insertion_2021,postava_anisotropy_2002,hamrle_-depth_2002} to the conventional magneto-optical Kerr effect (MOKE) signals. The complex Kerr angle at normal incidence for an infinitely thick film is given by 
$ \tilde\Theta_\mathrm{K} = \frac{ \tilde\epsilon_{xy}}{(1-\tilde\epsilon)}$, where $\tilde\epsilon_{xy}$ is the off diagonal component of the relative permittivity of the material \cite{you_generalized_1998}. We also define $\tilde S$ and $\tilde P$ which are sensitivity coefficients that depend on the (non-magneto-optical) optical properties of the magnetic layer: 
\begin{equation}
\tilde S=\frac{\tilde n(\theta_k)  }{\cos \theta_k\tilde n(\theta_k)+ \sin^2\theta_k} \label{Ssensitivity}
\end{equation}
\begin{equation}
\tilde P=\frac{\tilde n(\theta_k) }{\cos \theta_k \tilde n(\theta_k)- \sin^2\theta_k} \label{Psensitivity}
\end{equation}
From these definitions of $\tilde S$, $\tilde P$ and $ \tilde\Theta_\mathrm{K}$, the sources of the k-BLS signal expressed by $\tilde \Phi_s(z)$ and $\tilde \Phi_p(z)$ can be written as:

\begin{equation}
     \frac{\tilde {\Phi}_s(z)} {\tilde\Theta_\mathrm{K}\cos\theta_k }=  \tilde S \Big (\underbrace{\tilde m_{\Psi_k,{\mathrm{{z}}}} (z)}_\textrm{PMOKE} - \underbrace{{\tilde L \,} \tilde m_{\Psi_k,{\mathrm{{\phi_0}}}}(z)}_\textrm{LMOKE-like}   \Big) 
\label{phiS}  \end{equation}
and 
\begin{equation}  
    \frac{\tilde \Phi_p (z)}{ \tilde\Theta_\mathrm{K}\cos\theta_k }=  \tilde P \Big( \tilde m_{\Psi_k,{\mathrm{{z}}}} (z) + \tilde L \, \tilde m_{\Psi_k,{\mathrm{{\phi_0}}}} (z)  \Big) 
    \label{phiP} 
\end{equation}
where ${\tilde L}$ is the sensitivity to the longitudinal part of the dynamical magnetization:
\begin{equation}
    {\tilde L} = \frac{\sin(\phi_k - \phi_0)\sin\theta_k}{\tilde n(\theta_k)} \label{tildeL}
\end{equation}
with $\phi_0$ being the angle between the in-plane equilibrium magnetization and the $x$-axis.  
\subsection{Respective roles of the in-plane and out-of-plane components of the dynamic magnetization} 
The above expressions (Eq.~\ref{Ssensitivity}-\ref{phiP}) are instructive and therefore call for several comments. 

(a) The common scaling of $\tilde\Phi_s$ and $\tilde\Phi_p$ with $\cos\theta_k$ shows that the rays with grazing incidence ($\theta_k= \pi/2$) do not contribute to the overall $\mu$-BLS  signal. The additional scaling of the LMOKE contribution with $\sin\theta_k$ indicates that the sensitivity to the in-plane dynamic magnetization vanishes for normal incidence and is maximal for $\theta_k=\pi/4$. 

(b) The scaling of the longitudinal sensitivity $\tilde L$ with $\sin(\phi_k - \phi_0)$ in Eq.~\ref{phiS}  recalls that the dynamic magnetization contributes to the LMOKE part of the back-scattering BLS signal only if this dynamic part has a finite projection in the plane of incidence of the light, i.e. if $\phi_k \neq \phi_0$. 
\begin{table*}
\caption{Complex or real nature of the LMOKE-like contribution ${{\tilde L \,} \tilde m_{\Psi_k,{\mathrm{{\phi_0}}}}(z=0)}$ of the dynamical in-plane component of the magnetization to the sources of BLS signals (Eq.~\ref{phiS} and \ref{phiP}). By convention, the phase of the eigenvectors is taken so that $\tilde m_{\Psi_k,z}$ are real numbers at $z=0$. }
\begin{tabular}{|c|c|c|c|} 
    \hline
    \backslashbox{$\frac {1} {\tilde n(\theta_k)}$}{ellipticity} & &  $
    \tilde m_{\Psi_k,\phi_0} \in i\mathbb R^+$ & $\tilde m_{\Psi_k,\phi_0}^* \in i\mathbb R^-$\\ 
     &  & antiStokes, $\omega >0$   & Stokes, $\omega <0$\\ \hline
Drude metal: & &  ${{\tilde L \,} \tilde m_{\Psi_k,{\mathrm{{\phi_0}}}}(z=0)}$ &  ${{\tilde L \,} \tilde m_{\Psi_k,{\mathrm{{\phi_0}}}}^*(z=0)}$\\
     $ \in (1-i)\mathbb R^+$   & &  $\in (1+i)\mathbb R^+ \sin(\phi_k-\phi_0)$   &  $\in (1+i) \mathbb R^-\sin(\phi_k-\phi_0)$  \\ \hline
    Dielectric: & &  ${{\tilde L \,} \tilde m_{\Psi_k,{\mathrm{{\phi_0}}}}(z=0)}$ &  ${{\tilde L \,} \tilde m_{\Psi_k,{\mathrm{{\phi_0}}}}^*(z=0)}$\\
     $ \in \mathbb R^+$  & &  $\in i \mathbb R^+ \sin(\phi_k-\phi_0)$  &  $\in i \mathbb R^- \sin(\phi_k-\phi_0)$   \\ \hline
\end{tabular} 
\label{Lmoketerm}
\end{table*}

(c) Since $|\frac{\sin\theta_k}{\tilde n(\theta_k)}| \ll 1$ for most materials (especially for metals) and incidences, the scaling of the longitudinal sensitivity $\tilde L$ in Eq.~\ref{phiS} and Eq.~\ref{phiP} with the direction $\frac{\sin\theta_k}{\tilde n(\theta_k)}$ of the refracted pseudo-beam \footnote{Note that since at the surface of optically lossy materials like ours the in-plane component of the optical wavevector is real while its $z$ component is complex, there is no properly defined 'direction' of the refracted 'beam'. However since the optical wave obeys $\tilde k_\textrm{light}^2 = \tilde \epsilon \, \omega_\textrm{light}^2$, the refracted light for metallic films has a large $k_z$. This means that the refracted 'beam' can be qualitatively viewed as closer to the normal direction than the incoming beam. In practice for metallic films we can almost confuse $\tilde n(\theta_k) \approx \tilde n(\theta_k=0) $, 
see \cite{johnson_optical_1974}} is the reason why the sensitivity of the BLS signal to the in-plane longitudinal dynamic magnetization is much smaller than to the out-of-plane dynamic magnetization, even at large incidences \cite{hamrle_analytical_2010}\footnote{The transverse MOKE, appearing only when the incident light comprises some p-polarized part, does not alter the light polarization but generates a change in the $r_{pp}$ reflectivity \cite{buchmeier_intensity_2007,penfold_transverse_2002,martin_valderrama_layer-resolved_2024}. This effect is second order in dynamical magnetization and therefore can be neglected.}. This low sensitivity to the in-plane magnetization component will be illustrated in a specific example in the comparison of Figs.~\ref{calBLSPmokeLmoke}(a) versus (b).\\

(d) The last comment concerns the subtraction-versus-addition of the LMOKE signal from the PMOKE signal, which differentiates Eq.~\ref{phiS} from \ref{phiP}. Table~\ref{Lmoketerm} summarizes the effect of this subtraction-versus-addition onto the overall sources of BLS signal $\tilde {\Phi}_s$ and $\tilde {\Phi}_p$, depending on the optical properties and the sign of the frequency shift. Because spin waves are gyrotropic, the mode profiles all obey $\tilde m_{\Psi_k,{\mathrm{{\phi_0}}}}(z) / \tilde m_{\Psi_k,{\mathrm{{z}}}} (z) \in i \mathbb R^+$ in every slice of the material for $\omega >0$ (antiStokes processes) and $\tilde m_{\Psi_k,{\mathrm{{\phi_0}}}}^*(z) / \tilde m_{\Psi_k,{\mathrm{{z}}}}^* (z) \in i \mathbb R^-$ for $\omega <0$ (Stokes processes). \\

The importance of the optical properties can be seen by comparing the $\tilde L$'s of a perfect dielectric material (i.e. with $\tilde \epsilon \in \mathbb R^+$ and $\epsilon >1$) and that of a perfect Drude metal much below its plasmon frequency (enabling to assume $\tilde \epsilon \in i \mathbb R^+$ and $\textrm{Im}(\tilde \epsilon) \gg 1$. \\
For dielectric films, the LMOKE contribution ${{\tilde L \,} \tilde m_{\Psi_k,{\mathrm{{\phi_0}}}}(z)}$ underbraced in Eq.~\ref{phiS} is purely imaginary and small (Table~\ref{Lmoketerm}): it essentially rotates $\tilde {\Phi}_s(z)$ and $\tilde {\Phi}_p(z)$ in opposite directions in the complex plane. As a result, $\tilde {\Phi}_s(z)$ and $\tilde {\Phi}_p(z)$ have amplitudes that differ only thanks to the tiny difference between $\tilde S$ and $\tilde P$. The arguments of $\tilde {\Phi}_s(z)$ and $\tilde {\Phi}_p(z)$ differ thanks to the longitudinal contribution. \\ 
In contrast, metallic materials have near imaginary diagonal permittivity $\tilde \epsilon$, so that $\textrm{Im}[\frac 1 {\tilde n(\theta_k)}] \approx - \textrm{Re}[\frac 1 {\tilde n(\theta_k)}] < 0 $. As a result, the LMOKE contribution ${{\tilde L \,} \tilde m_{\Psi_k,{\mathrm{{\phi_0}}}}(z)}$ underbraced in Eq.~\ref{phiS} rotates them in opposite directions of the complex plane but \textit{also reduces }the modulus of $\tilde {\Phi}_s(z)$ and \textit{increases} the modulus of $\tilde {\Phi}_p(z)$. Moreover we have $|\tilde S| < |\tilde P|$: this explains why k-BLS measurements performed with a s-polarized light give less intensity than when performed with p-polarized light.

\subsection{\label{SASk-resolv} Relation between non-reciprocity and Stokes/Anti-Stokes asymmetry}

Both $\mu$-BLS and k-BLS spectra generally exhibit some S/AS asymmetry, even at room temperature when magnons can be considered as classical spin waves with essentially equal (and very large) populations after annihilation or after the creation of a single SW. The above model can identify three reasons \footnote{Note that as considered so far, we continue to assume that the quadratic magneto-optical effects can be neglected.} for the S/AS asymmetry of k-BLS spectra, arising respectively (i) purely from the SW non-reciprocity (NR), (ii) from a combination of optical properties and NR of the SW mode profiles within the depth of the material, and (iii) finally from sole optical absorption in the case of perfectly reciprocal materials.  
\subsubsection{Case of frequency non-reciprocity}
Many magnetic systems possess frequency non-reciprocity (i.e. $\omega_{\Psi_k} \neq \omega_{\Psi_{-k}}$). This is for instance occurring in single layers with interface DMI \cite{belmeguenai_interfacial_2015}, in single layers with different anisotropies at the two interfaces \cite{gladii_frequency_2016}, or in multilayers as a result of dipole-dipole interaction \cite{stamps_spin_1994, verba_wide-band_2019}. 
The modes $\Psi_k$ and $\Psi_{-k}$ are affected respectively by antiStokes and Stokes processes. The k-resolved BLS spectrum is thus formed by signals of the type: $$\eta(\Psi_k)\ I(\Psi_k) \,\delta(\omega-\omega_{\Psi_k}) + \eta(\Psi_{-k})\ I(\Psi_{-k}) \,\delta(\omega+\omega_{\Psi_{-k}}).$$where $\delta$ is the usual Dirac function or its finite linewidth equivalent (Eq.~\ref{lorentz}), summed over the different spin wave branches. Since $\omega_{\Psi_k} \neq \omega_{\Psi_{-k}}$, there will be a S/AS asymmetry: the frequencies of the peaks within the positive and negative frequency parts within the k-BLS spectra differ. 
\subsubsection{Case of non-reciprocity of the mode profiles}
The second case is when there is frequency reciprocity ($\omega_{\Psi_k} = \omega_{\Psi_{-k}}$) but the modes have a non-reciprocal profile, i.e. $\tilde m(z,\Psi_k) \neq \tilde m(z,\Psi_{-k})$. This is for instance occurring in Damon-Eshbach (DE) spin waves that tend to localize near the top or bottom interface of a film, depending on the sign of the wavevector \cite{damon_magnetostatic_1961, camley_stokesanti-stokes_1982}. In this case, we can have equal thermal populations of the modes (i.e. $\eta(\Psi_k)= \eta(\Psi_{-k})$) and the peaks within the BLS spectra will be at the same negative and positive frequencies. However, since light is absorbed and dephased upon its travel through the depth of the magnetic material (as described by the factor $e^{\frac{i 4 \pi \tilde n z}{\lambda}}$ in Eq.~\ref{electricfieldps}), the difference in depth dependence of the dynamic magnetization leads to $E^{\mathrm{BLS}}_{p, s}{(\Psi_k)} \neq E^{\mathrm{BLS}}_{p, s}{(\Psi_{-k})}) $ in Eq.~\ref{electricfieldps} and therefore $I(\Psi_k)\neq I(\Psi_{-k})$. In this case, the combination of light absorption and non-reciprocity of the mode depth profile leads to a S/AS asymmetry of the amplitudes of the peaks at negative versus positive frequencies within k-BLS spectra. 
\subsubsection{Case of perfect reciprocity}
Let us finally consider the case of perfect reciprocity, i.e. 
$\omega_{\Psi_k} = \omega_{\Psi_{-k}} $ and $\tilde m(z,k) = \tilde m(z,-k)$.  Even in these conditions, the k-BLS spectra generally exhibit S/AS asymmetry \cite{sandercock_light_1978,camley_stokesanti-stokes_1982,grunberg_light_1977}. The S/AS symmetry is recovered only when the incoming light is purely $s$-polarized ($E_p=0$) and an incidence such that $\phi_k=\phi_0$ is used \cite{zivieri_stokesanti-stokes_2002}. In the general case, the S/AS asymmetry for perfectly reciprocal materials comes from optical effects that affect the relative phases of the scattering contributions of the in-plane and out-of-plane dynamic components of the magnetization. This can be understood from the above formalism. \\ 
As can be indeed seen in Table~\ref{Lmoketerm}, the $ \tilde L \tilde m_{\Psi_k,{\mathrm{{\phi_0}}}}$ and $\tilde m_{\Psi_k,z}$ (complex-numbered) contributions to the scattered light 
add in the antiStokes part of the spectrum, while they subtract in the Stokes part of the spectrum. The add-versus-subtract situation is reversed when considering the contribution of the p-polarized light instead of the s-polarized light (see Eqs.~\ref{phiS}-\ref{phiP}). For arbitrary polarization of the incident beam and an optically absorbing medium, the combination of Eqs.~\ref{EpEs} and Eq. \ref{IdePhSIk} thus leads to some S/AS asymmetry. The symmetry is recovered only when the incoming light is purely $s$-polarized ($E_p=0$) and an incidence such that $\phi_k=\phi_0$ is used, thereby canceling the influence LMOKE contribution (see Zivieri et al. in \onlinecite{zivieri_stokesanti-stokes_2002}). 

\subsubsection{Consequence for microfocused BLS spectra}
The $\mu$-BLS signal $F_{bls}$ is thus a sum of k-BLS spectra that are asymmetric for most of them.
This sum should most often result in asymmetric $\mu$-BLS spectra (i.e. $F_{\mathrm{bls}}(\omega) \neq F_{\mathrm{bls}}(-\omega)$). We have indeed confirmed experimentally such S/AS asymmetry of $\mu$-BLS spectra on BiYIG samples (not shown), CoFeB samples (Fig.~1) and on CoFeB/Ru/CoFeB synthetic antiferromagnet samples (here the cancellations of the dynamic magnetizations of the two layers lead to S/AS intensity ratios that are close to --but distinct from-- unity, not shown).

\section{\label{sec:CoFeBapp}Application to a single layer}
This section implements the model on a 50 nm thick $\mathrm{Co}_{40}\mathrm{Fe}_{40}\mathrm{B}_{20}$ magnetic layer, whose properties are listed in Table \ref{materialparam}.  The laser polarization, the static field and the equilibrium magnetization are oriented along the $x$ axis. A prerequisite for the implementation of the microfocused BLS model is to identify the spin wave eigenmodes $\Psi_k$ of the system, their frequencies and their linewidths. Although not needed for the calculation of the BLS spectra, the density of SW states (DOS) will be useful for their understanding.

\subsection{Dispersion relations}
\begin{table}[]
    \centering
    \caption{Values of the inputs used for calculating the spin wave properties and the optical response at a wavelength of 532 nm. The daggers indicate that the values were measured by VNA-FMR. The double dagger indicates that the values were measured by ellipsometry assuming a semi-infinite homogeneous film.}
\begin{tabular}{|m{4.0cm}|m{3.8cm}|} 
\hline
  Definitions & Values \\   
\hline
   Static field ($\parallel \vec e_x$) & 30 mT \\   
   Magnetization & 1.77 T \\   
   Exchange stiffness  & 16 pJ/m \\   
   Gilbert damping & 0.004~$\dagger$ \\      
   Thickness & 50 nm \\  
   Gyromagnetic ratio & 29 GHz/T~$\dagger$ \\     
   Effective diagonal optical permittivity $\tilde \epsilon$ of the sample & $-9.97 + 14.93i$~$\dagger\dagger$ \\     
   Off-diagonal optical permittivity $\tilde \epsilon_{xy}$ of Co~\cite{fu_dielectric_1995} & $0.435 - 0.107i$ \\     
   Resulting Kerr angle at normal incidence $\tilde \Theta_\textrm{Kerr}$ & $0.0186 + 0.0155i$ (radians) \\    
   Skin depth & $\lambda / (4 \pi \mathrm{Im}(\tilde n)) \approx 11.3$ nm \\ 
   Phase rotation length & $\lambda / (4 \pi \mathrm{Re}(\tilde n)) \approx 21.2$ nm \\ 
   Experimental spectral resolution & $\xi = 570$ MHz \\ 
\hline
\end{tabular}
\label{materialparam}
\end{table}

\begin{figure*}
    \centering
    \includegraphics[scale=0.1,width =17cm]{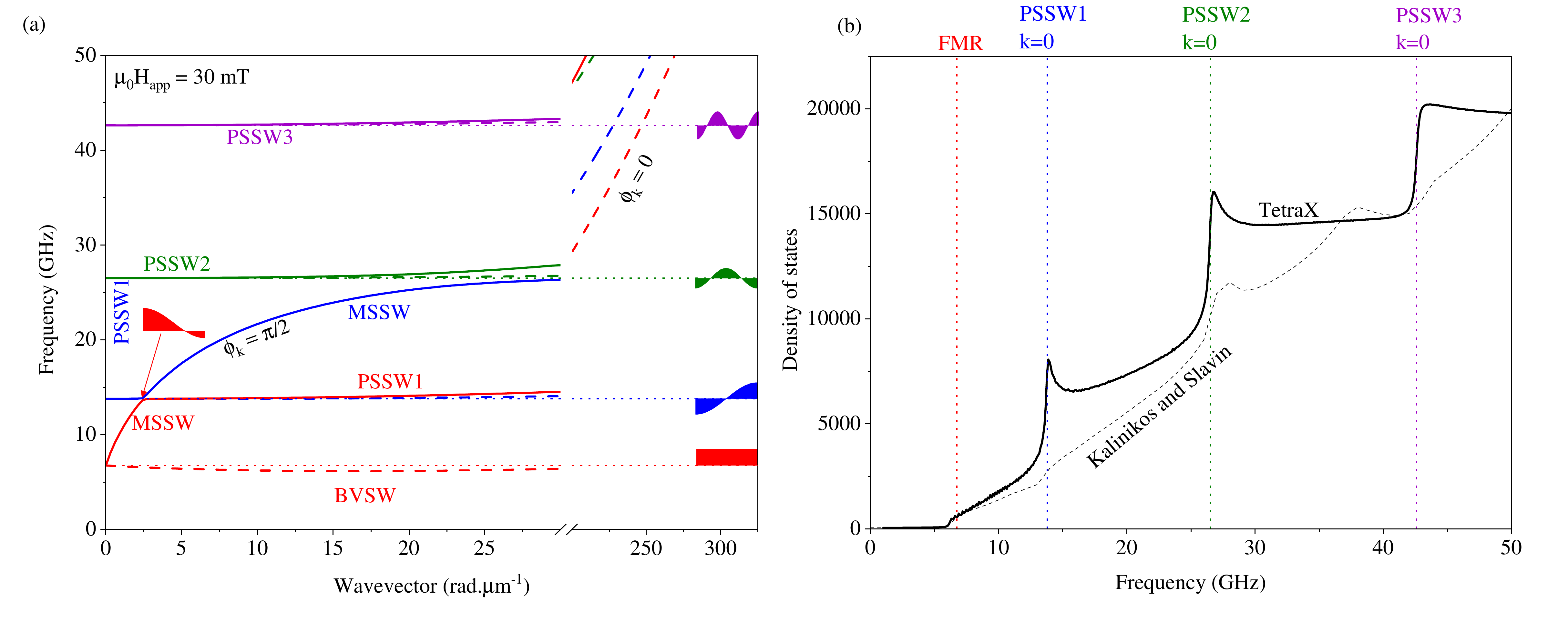}
    \caption{Calculated spin wave properties for a field of 30 mT using the parameters Tab.~\ref{materialparam}. (a) Spin wave dispersion relations according to TetraX in the Damon-Eshbach geometry (bold lines, $\vec k \perp \vec m_0$) and the backward volume geometry (dashed lines, $\vec k \parallel \vec m_0$). The dotted lines are the frequencies at $\vec k=\vec 0$. Insets: mode profiles $\textrm{Re}\big(\tilde m_{\phi_0} (z) \big)$ of the FMR, PSSW1,PSSW2 and PSSW3 modes at $\vec k=\vec 0$, and of the MSSW/PSSW1 mode at $k_y=2.5~\textrm{rad}/\mu \textrm{m}$ in the region where there is hybridization between the MSSW wave and the first perpendicular standing spin wave. (b) Resulting density of spectral functions of spin waves according to TetraX (bold line) and according to the Kalinikos and Slavin model (\onlinecite{kalinikos_theory_1986}, thin dotted black line).}
    \label{tetraxouput}

\end{figure*}
We use the TetraX package \cite{korber_finite-element_2021,korber_numerical_2021} to numerically obtain \footnote{The sample thickness is meshed in 1 nm-thick sheets. } the spin wave frequencies, their linewidths and their profiles for all $||\vec k||$ up to $300~\textrm{rad}/\mu \textrm{m}$. Such a high value exceeds by far the need for the forthcoming BLS modeling but it is needed to properly account for the DOS up to 50 GHz. Figure \ref{tetraxouput} (a) plots the dispersion relations of the backward-volume magnetostatic spin 
wave (BVSW), magnetostatic surface spin waves (MSSW) and the three PSSW modes that exist below 50 GHz. To better reveal that the dispersion relations are anisotropic, we plotted them for two wave orientations: the classical "backward volume" configuration (BVSW, $\vec k \parallel H_\textrm{app}$, $\phi_k = 0$) and the "Damon-Eshbach" configuration (DE, $\vec k \perp H_\textrm{app}$, $\phi_k = \pi/2$). 

The insets in the right part of Fig.~\ref{tetraxouput} (a) display the thickness profiles of the modes at $ \vec k$ = 0. As expected, the lowest frequency mode is the FMR mode presenting a uniform amplitude, and the three higher-frequency modes are perpendicular standing spin waves (PSSW) with cosine profiles exhibiting $p=1,2,3$ nodes. In the BV configuration, the 4 modes are almost non-dispersive with almost flat $\omega(k)$'s. In contrast, the modes --especially the two lowest frequency ones-- are clearly dispersive in the DE configuration. The MSSW and PSSW1 anticross at $|\vec k| \approx$ 2.5 rad.$\mu$$\mathrm{m}^{-1}$. This anticrossing reveals the hybridization of the two modes \footnote{Note that this feature is not included in the Kalinikos-Slavin theory which does not account for the dipolar interaction between the modes}, which is particularly clear when looking at the mode profile [see the red arrow in Fig.~\ref{tetraxouput} (a)]. This profile presents indeed a higher amplitude near one interface --which is reminiscent of a "DE" character \cite{kalinikos_theory_1986}-- and a node within the thickness --which is reminiscent of the first perpendicular spin wave mode character--.

\subsection{Density of states}
The dispersion relations can be used to calculate the density of spectral functions (DOS):
\begin{equation}
    \mathrm{DOS}(\omega)=\sum_{\Psi_{k}} \mathcal L_{\Psi_k} (\omega) 
    \label{dps}
\end{equation}

The sum runs over a square grid of $\vec k$ values with steps of 1 rad.$\mu$$\mathrm{m}^{-1}$ in both $k_x$ and $k_y$ directions.
Fig.~\ref{tetraxouput}(b) plots the DOS. It vanishes under a frequency slightly below the FMR frequency: this corresponds to the absolute minimum of the dispersion, which happens for the BV configuration at $k_\textrm{min}=16.3~\textrm{rad}/\mu$m and $\omega/(2\pi)=$ 6.15 GHz. The density of states peaks at the PSSW1, PSSW2 and PSSW3 frequencies, where Van Hove singularities occur because of the vanishing dispersions at $ \vec k=\vec 0$ for these three modes. In contrast, the DOS does not peak in the frequency region of FMR: the Van Hove singularity is avoided because the minima of the dispersion relations are distributed from $|k|=0$  for $\phi_k=\pi/2$ to $k_\textrm{min}$ for $\phi_k=0$. Fig.~\ref{tetraxouput} (b) plots also the DOS that would be calculated from Kalinikos and Slavin theory \cite{kalinikos_theory_1986}; as noted by the authors themselves, the dipolar interactions between the modes are not taken into account in this model and the explicit (but simplified) form of their dispersion relations are valid only up to the mode crossing. As a result, this theory overestimates the dispersive aspect of the spin waves \footnote{See supplementary material for further details}, which leads to an underestimation of the DOS and a softening of the Van Hove singularities. Thus, all forthcoming evaluations will rely on the SW properties calculated from the TetraX package.

\begin{figure}
    \centering
    \includegraphics[width =8.2cm]{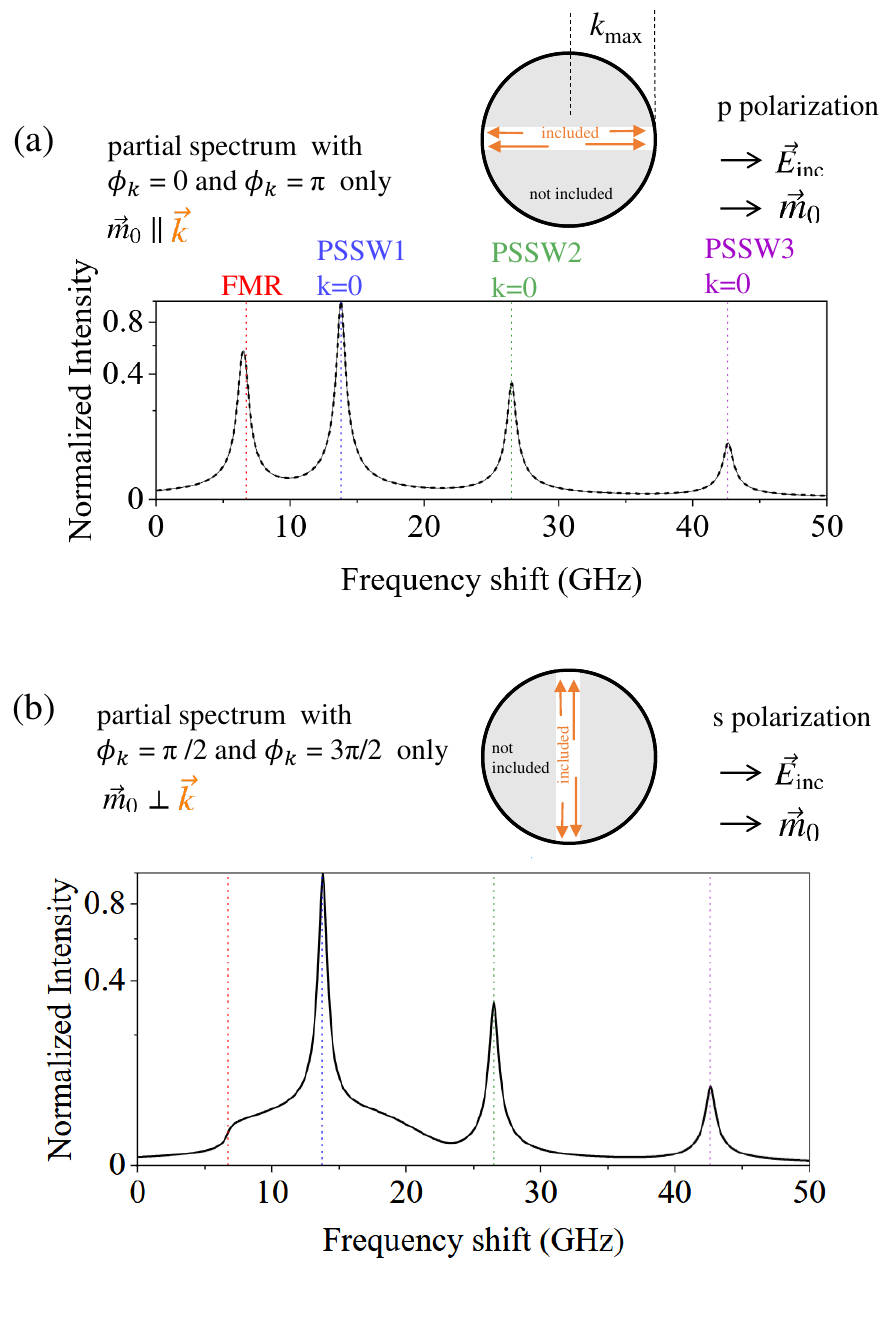}
    \caption{
    Comparative contributions of the spin waves of selected wavevector orientations. Partial $\mu$-BLS antiStokes spectra obtained if the only spin waves populated in the material are (a) the (almost non-dispersive) backward volume waves ($\phi_k$=0 and $\phi_k$=$\pi$ ), corresponding to the situation of Fig.~\ref{shcmconvt}(a) or (b) the (more dispersive) Damon-Eshbach waves ($\phi_k$=$ \frac{\pi}{2}$ and $\phi_k$=$\frac{3\pi}{2}$), corresponding to Fig.~\ref{shcmconvt}(b). 
    The vertical dotted lines correspond to the frequencies of the FMR mode and the PSSW 1,2 and 3 modes. Each spectrum is normalized with respect to the $p = 1$ peak. The material properties are listed in Table~\ref{materialparam}. The spectral intensities are scaled in a square root manner to better evidence the low intensity features}.
    \label{calBLSDEBV}
\end{figure}

\subsection{Relation between microfocused BLS spectra and density of states} \label{sectionSpectraVersusDOS}
The microfocused BLS spectra and the density of states both contain contributions of \textit{all} spin waves present in the system. The BLS spectra and the DOS should therefore share common features. The common features are conveniently revealed by performing the two fictitious BLS experiments displayed in Fig.~\ref{calBLSDEBV}. 
The first fictitious experiment considers that the $\vec k \parallel \vec e_x \parallel \vec m_0$ (abusively, the "Backward Volume", BV) waves are the sole to be populated in the system [Fig.~\ref{calBLSDEBV}(a)] and the other that the $\vec k \parallel \vec e_y \perp \vec m_0$ (abusively, the "Damon-Eshbach", DE) waves are the sole populated in the system [Fig.~\ref{calBLSDEBV}(b)]. This is done by restricting the sum in Eq.~\ref{BLSspectrumMasterEquation} to the corresponding orthogonal line scans in wavevector space. \footnote{Note that since the light polarization is $\vec E_\textrm{inc} \parallel \vec e_x$, the BV waves are thus sensed with a p-polarization, while the DE waves are sensed with an s-polarization. This difference has little impact on the conclusions of section \ref{sectionSpectraVersusDOS}.}
\subsubsection{Spectral contribution of the sole waves with wavevectors along x: $\mu$-BLS peaks for non-dispersive modes}
The BVSWs are almost non-dispersive [see Fig.~\ref{tetraxouput}(a)] such that the partial DOS corresponding to these modes only (not shown) consists of 4 discrete narrow peaks of identical areas, as if these modes were the discrete energy levels of an atom.
If these sole BVSWs were populated, the $\mu$-BLS spectrum [Fig.~\ref{calBLSDEBV}(a)] would contain 4 narrow peaks positioned at frequencies very near the frequencies of the FMR and the PSSW 1, 2 and 3 modes. 

Notably, the $p=1$ peak gives the most intense peak in the spectrum [Fig.~\ref{calBLSDEBV}(a)]. This could seem counterintuitive at first sight, because owing to their lower frequencies, the thermal populations $\eta_k$ of the BVSW modes responsible for the $p=0$ peak should be approximately twice that of the PSSW1 modes responsible for the $p=1$ peak, such that twice more magnons contribute to the area of the $p=0$ peaks in the BVSW-only BLS spectrum. 
The overlap between the electromagnetic field profile of the optical waves within the material and the magnetization profile within the material cannot explain the relatively weak $p= 0$ peak (see the supplementary material for a discussion of the respective roles of these two profiles): 

The comparatively low signal amplitude of the $p=0$ peak can rather be understood from the ellipticities of the involved modes. The BVSW (PSSW1) modes have a large (low) ellipticity. The out-of plane $\tilde{m}_z$ components of the BVSW quanta are much smaller than that of the PSSW quanta, whose precession is much more circular. As the BLS signal of BVSW in the p-polarized light case of Fig.~\ref{calBLSDEBV}(a) is purely coming from the PMOKE-like contribution, this contrast of the $\tilde{m}_z$'s counterbalances the contrast of the $\eta_k$'s, yielding a $p=0$ peak that is smaller than the $p=1$ peaks.

\subsubsection{Spectral contribution of the sole waves with wavevectors along y: smeared spectrum for dispersive modes}
The situation is very different when instead of looking at the spectrum constructed from the sole backward volume spin waves, we now consider the contribution of the sole "Damon-Eshbach" spin waves [Fig.~\ref{calBLSDEBV} (b)], because the latter are very dispersive [see Fig.~\ref{tetraxouput}(a)]. Only the three high-frequency peaks remain in the spectrum. The contributions of the very dispersive two lowest spin wave branches (MSSW and PSSW1 modes, as well as their hybrids at the anticrossing) spread over a wide frequency interval (6-25 GHz), giving a small asymmetric shallow hill skewed to the high frequency side. The frequency width of this shallow hill has nothing to do with the Gilbert linewidth of the modes and with the spectral resolution $\xi$ of the setup. It extends typically up to the frequency of the MSSW mode at the highest wavevector accessible by the BLS setup, i.e.:
\begin{equation}
\frac{\omega}{2 \pi}\big(\textrm{MSSW}\,\phi_k=\frac \pi 2,\,k_\mathrm{max}=18~\textrm{rad}/\mu\textrm{m}\big) \approx 25\,\textrm{GHz}. \label{25GHz}
\end{equation}
\subsubsection{Spectral contribution of the full set of spin waves}
\begin{figure}
    \centering
    \includegraphics[scale=0.1,width =8cm]{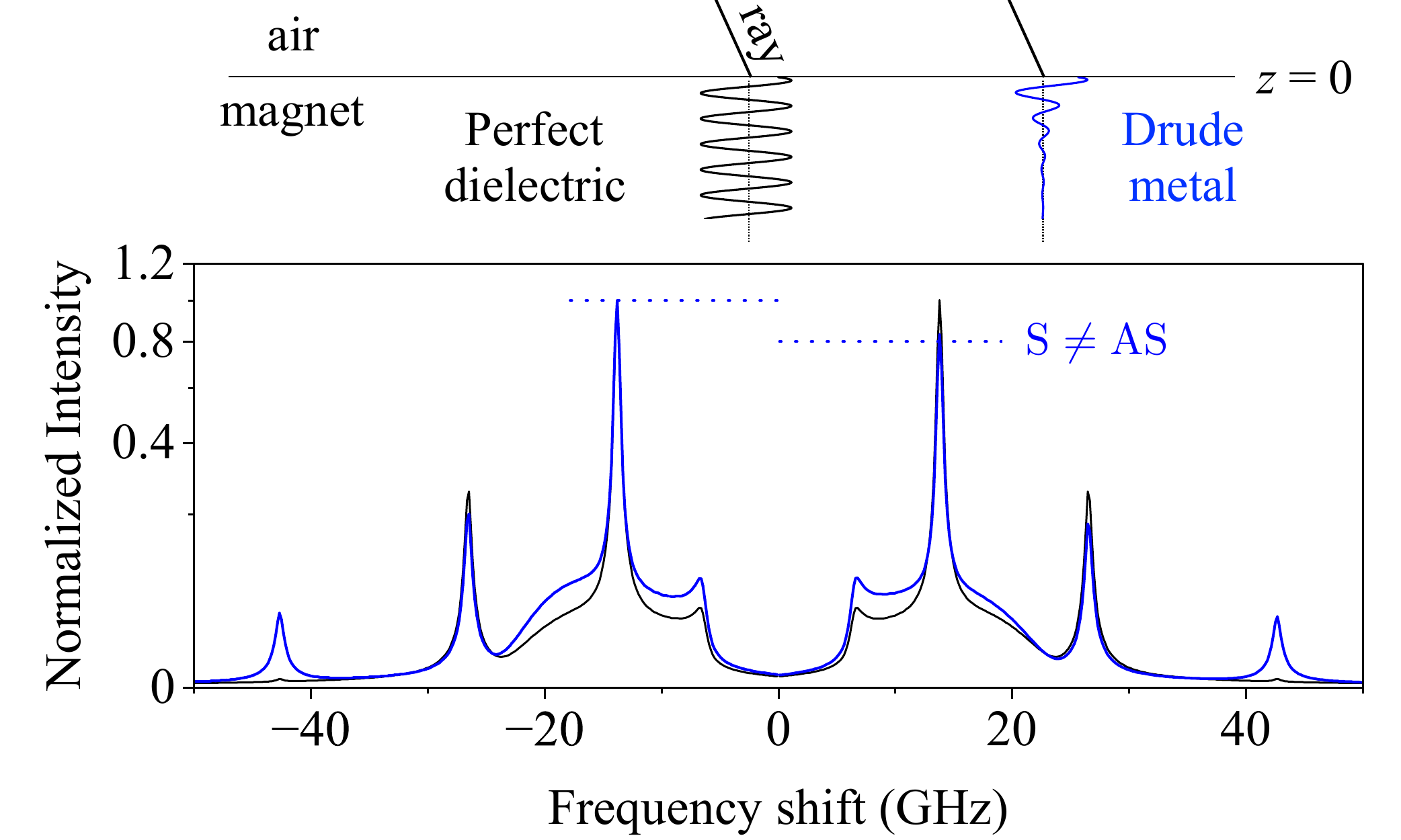} 
    \caption{Influence of the optical properties. Top panel: sketch of the light penetration when the sample is a perfect dielectric (black line) or a Drude metal (blue line). Bottom panel: Modeled $\mu$-BLS spectra that would be obtained on a sample having the magnetic properties of CoFeB but with a permittivity of $\tilde \epsilon = 10$ (black spectrum), or with $\tilde \epsilon = 10 i $ (blue spectrum), the latter exhibiting Stokes/antiStokes asymmetry. 
    The magnetic properties used for this figure are listed in Table~\ref{materialparam}. \label{DielectricVersusDrudeMetal}}

\end{figure}

The $\mu$-BLS spectrum calculated for the two extreme geometries helps us to interpret the peak shapes observed in the $\mu$-BLS spectrum.  From this analysis, we can anticipate several conclusions: \\
(i) The shape of the $p=0$ peak is mainly determined by the large dispersion of the $\omega(\vec k)$'s and the populations $\eta(\vec k)$ of its contributing modes, with some influence of interplay between the thickness profile of their dynamic magnetization and the penetration of the light beam.
The rather small amplitude and composite shape of the $p=0$ peak observed experimentally [Fig.~\ref{resultscofeb}(a)] result from the sum of a rather narrow peak approximately placed at the absolute bottom of the dispersion relation, plus a shallow asymmetric hill offset to larger frequencies and extending up to the frequency of Eq.~\ref{25GHz}. The $p=1$ peak emerges from the shallow hill of the $p=0$ family of modes. \\
(ii) In contrast, the $p \geq 2$ peaks are much narrower since their $\omega(\vec k)$'s are essentially non-dispersive in the relevant range of wavevectors. The amplitudes of the $p\geq 2$ peaks are determined by their thermal populations --approximately inversely proportional to their frequencies-- and the interplay between their thickness profile and the light penetration profile within the magnetic material. This phenomenon is bound to depend on the optical properties of the sample, with an expected difference between metallic and dielectric samples, as will be illustrated in the next section and in the example of Fig.~\ref{DielectricVersusDrudeMetal}. \\
(iii) In all cases, the linewidth of each peak in a $\mu$-BLS  spectrum is unlikely to be informative of the Gilbert lifetime of the spin waves. It reflects the dispersion of the spin waves within the interval of wavevectors sensed by the microscope objective, as well as the spectral resolution of the setup.

\subsection{Dependence of the microfocused BLS  spectra on the optical parameters: metals versus dielectrics}

We have conjectured that the light penetration profile determines the respective contributions of the in-plane and out-of-plane dynamical magnetization components, with consequences on the amplitude of the BLS response of the different modes, as well as an impact on the S/AS asymmetry.

To best reveal this importance of the optical properties, we calculated in Fig.~\ref{DielectricVersusDrudeMetal} the $\mu$-BLS spectra for two textbook materials: first a perfect dielectric (i.e. with $\tilde \epsilon \in \mathbb R^+$ and $\epsilon >1$) with the magnetic properties of CoFeB, then a perfect Drude metal much below its plasmon frequency (enabling to assume $\tilde \epsilon \in i \mathbb R^+$ and $\textrm{Im}(\tilde \epsilon) \gg 1$ at the wavelength of the BLS laser \cite{ordal_optical_1983}) with the same magnetic properties. 

When the material is metallic (in fact, as soon as it is not perfectly transparent), the antiStokes side of the spectrum is reduced while that of the Stokes side is increased, which induces some S/AS asymmetry. This shows that generally speaking, the amplitude of the peaks within an $\mu$-BLS spectrum can thus be interpreted only if accounting for the optical properties of the sample.

It is worth noticing that the $p=3$ peak almost disappears in the dielectric case (Fig.~\ref{DielectricVersusDrudeMetal}), in contrast to the metallic case. This quasi-disappearance is due to the perfectly odd thickness profile of the PSSW3 modes responsible for the $p=3$ peak, the odd character being maintained in the full range of accessible wavevectors. We stress that the PSSW1 modes, because of their hybridized character with the MSSW wave, do not have an odd thickness profile, and therefore the $p=1$ peak will be measurable irrespective of the material optical properties.

\subsection{Dependence of the microfocused BLS spectra on the ellipticity of the precession: Stokes/antiStokes asymmetry}

\begin{figure}
    \centering
    \includegraphics[width =8cm]{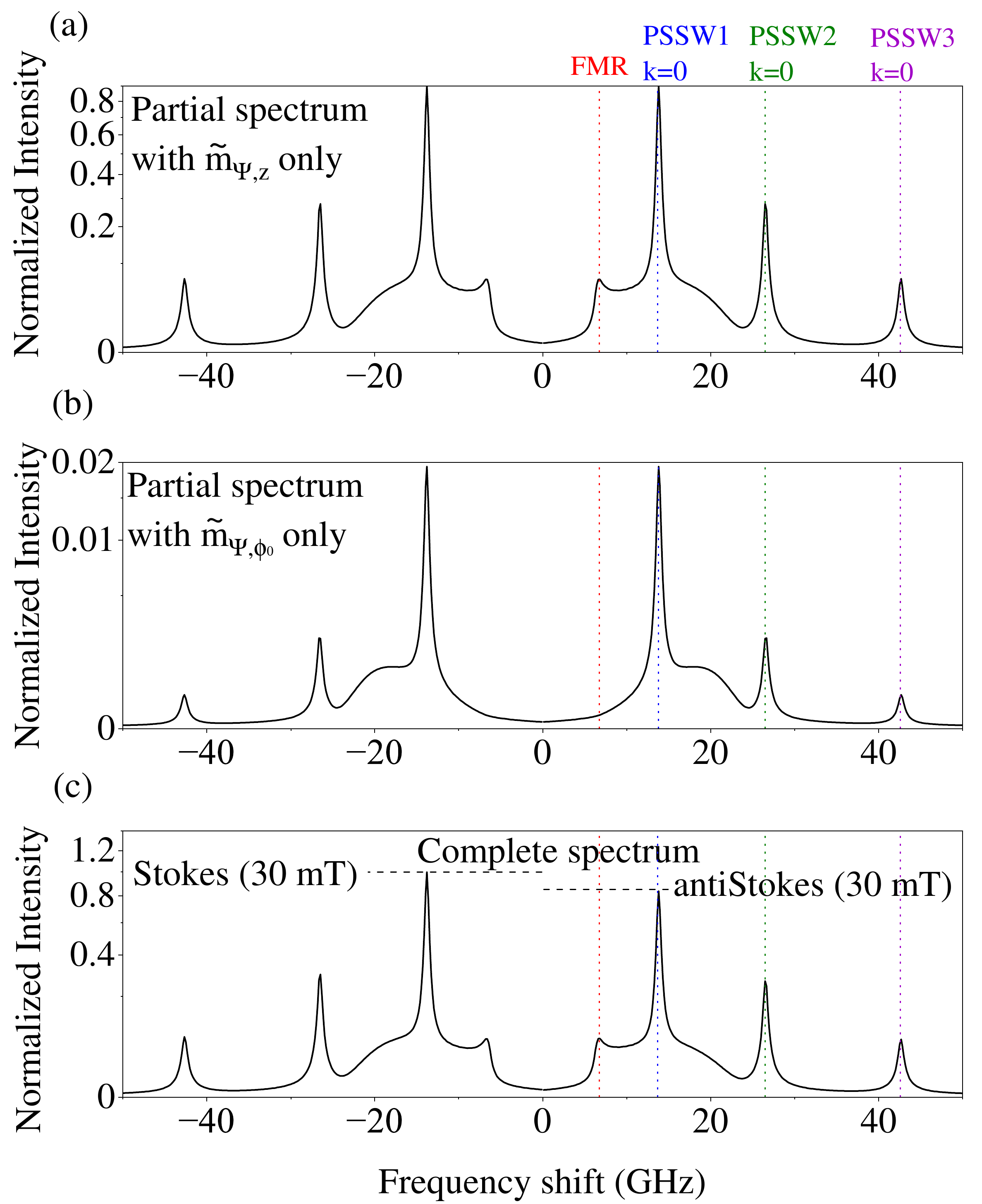}
    \caption{Effect of the ellipticity of the precession: modeled $\mu$-BLS 
spectra that would be obtained if in Eqs.~\ref{phiS} and \ref{phiP} (a) we only considered the out-of-plane dynamical component of the magnetization $\tilde m_{\Psi_k,{\mathrm{{z}}}}$, or (b) if we only considered the in-plane dynamical component of the magnetization $\tilde m_{\Psi_k,\phi_0}$, or (c) if the complete dynamical magnetization was taken into account. 
The spectra are normalized to the Stokes p=1 peak of panel. The optical and magnetic properties used for the calculations of this figure are listed in Table~\ref{materialparam}. The spectral intensities are scaled in a square root manner to better evidence the low intensity features. 
}

    \label{calBLSPmokeLmoke}
\end{figure}
As mentioned in the modeling section, the in-plane $\tilde m_{\Psi_k, \phi0}$ and out-of-plane $\tilde m_{\Psi_k, z}$ dynamical components of the magnetization contribute to the $\mu$-BLS  with weights that depend on the optical parameters. To separate their contribution to the $\mu$-BLS  signal, we calculate in Fig.~\ref{calBLSPmokeLmoke} the $\mu$-BLS  spectra corresponding to the two additional unphysical --but pedagogically instructive-- situations, where we successively set the precession ellipticity to linear polarization along the two axes of the dynamical magnetization. In panel (a), we set that $\forall \Psi_k$, $\forall z$, $\tilde m_{\Psi_k, \phi0}=0$. In panel (b) we set that $\forall \Psi_k$, $\forall z$, $\tilde m_{\Psi_k, z}=0$. The last panel of Fig.~\ref{calBLSPmokeLmoke} reports the $\mu$-BLS  spectrum of the real situation, i.e. obtained when considering the two dynamical components of all spin waves.

The first two spectra --the unphysical situations-- both lead to S/AS symmetry. The signal is large for the PMOKE-only spectrum  [Fig.~\ref{calBLSPmokeLmoke}(a), infinite precession ellipticity leading to $\tilde m_{\Psi_k, \phi_0}=0$] and comparatively much lower for the LMOKE-only spectrum [panel (b), opposite infinite precession ellipticity leading to $\tilde m_{\Psi_k, z}=0$]. The accounting of the two dynamical components of the spin waves (or in other words, the PMOKE and the LMOKE parts of the signal) restores the S/AS asymmetry: the LMOKE and PMOKE contributions interfere constructively on the Stokes side, and partially destructively on the antiStokes side, as was anticipated from Table \ref{Lmoketerm}.

The Fig.~\ref{calBLSPmokeLmoke}(b) deserves a last comment, notably because artificially suppressing the z component of the dynamic magnetization suppresses the $\mu$-BLS signal in the frequency region of FMR. The reason lies in Eq.~\ref{tildeL}: since $\tilde L \propto \sin \theta_k$, the LMOKE contribution to the BLS signal is negligible in the $k \approx 0$ region, such that the bottom of the $p=0$ band does not contribute to the LMOKE-part of the BLS signal.

\subsection{Comparison with experiment}

\begin{figure*}
    \centering
    \includegraphics[width =12cm]{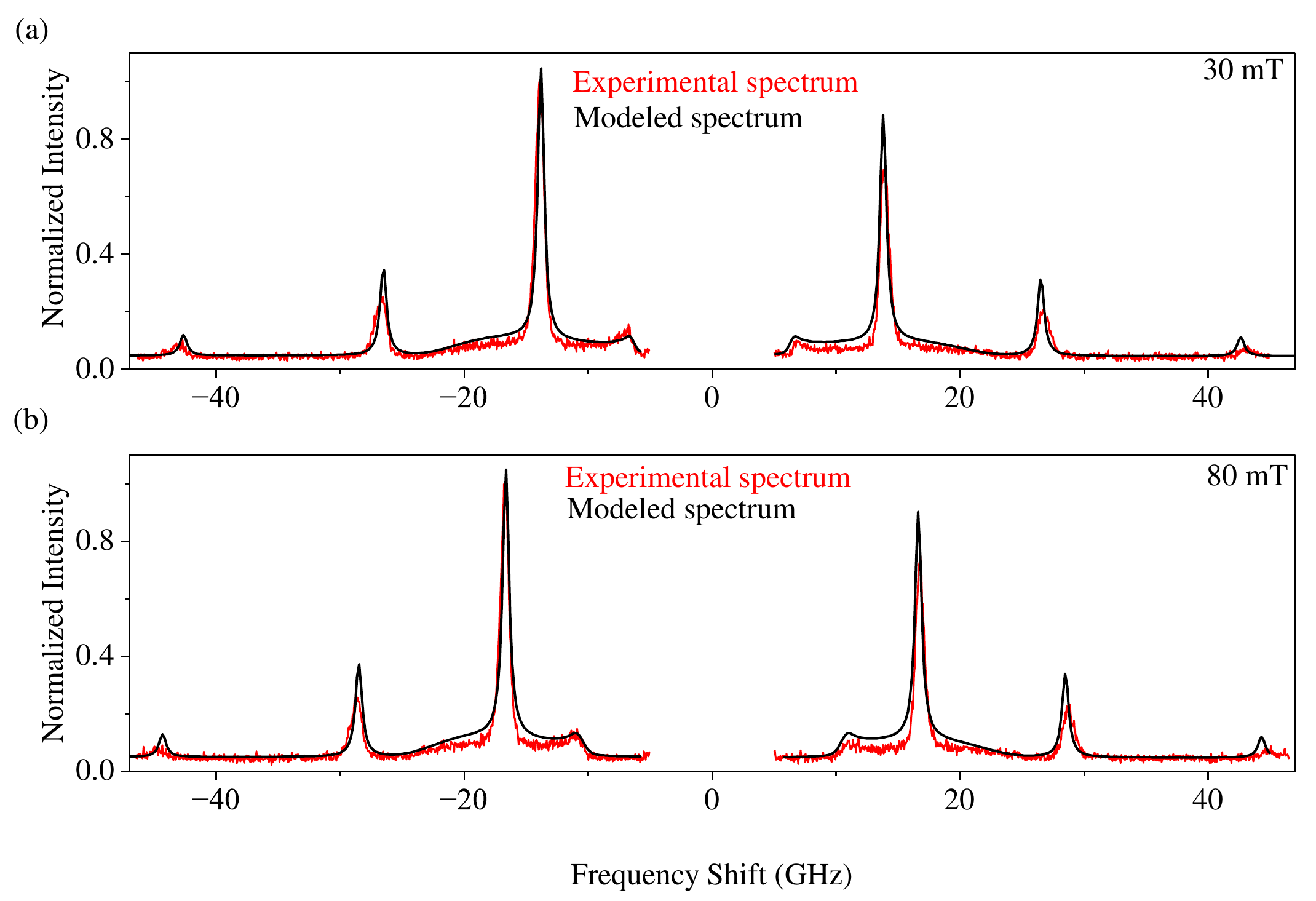} 
    \caption{Comparison between the experimental $\mu$-BLS  spectra (red lines) and our model (black lines). The experimentally determined noise floor was added to the theoretical spectra. (a) Spectrum recorded for an applied field of 30 mT. (b) Idem for 80 mT. The magnetic and optical properties used for this figure are that of Table~\ref{materialparam}. \label{expvsmodel3080mT}}

\end{figure*}

The $\mu$-BLS  spectra calculated from Eq.~\ref{BLSspectrumMasterEquation} using the parameters from Table~\ref{materialparam} are compared to the corresponding experimental result in Fig.~\ref{expvsmodel3080mT}. The modeled and experimental frequencies of the peaks match for these specific two applied fields, as well as for all fields listed in Fig.~\ref{resultscofeb} (not shown). This indicates that the magnetization, the gyromagnetic factor and the ratio $A_\textrm{ex}/t^2$ in Table \ref{materialparam} are correct. Since the thickness $t$ is precisely known, the exchange stiffness $A_\textrm{ex}=16$ pJ/m can be regarded as correct from the sole criterion of these four frequencies. 
Note however that other pairs $\{A_\textrm{ex}', t'\}$ with the same ratio $A_\textrm{ex}/t^2$ would lead to the same $p=0-3$ frequencies at $\vec k=\vec 0$. As a result, the $\{A_\textrm{ex}', t'\} \neq \{A_\textrm{ex}, t\}$ pairs lead to altered dispersion relations and the shape of the corresponding spectra are substantially modified and no longer match with the experimental result (see supplementary material).

In contrast, with the material parameters of Table \ref{materialparam}, the model optimally reproduces the experimentally observed shapes of the different spectral peaks, their intensities, and their S/AS intensity ratios. This strengthens our assumption that the optical properties of our multilayered material can be approximated as that of a single layer of semi-infinite thickness. 

Though satisfactory, the agreement between experiments and theory is not perfect, especially for the amplitude of the smeared $p=0$ and $p=1$ peaks. 
This remaining disagreement may result from our neglecting the skew scattering contributions to $\mu$-BLS spectra. The quantitative accounting of skew scattering is beyond the scope of this paper, but its effect can be qualitatively evaluated. 

In the appendix, we show that the existence of the skew scattering (in addition to the back-scattering) tends to even further mix the contributions of the different SW wavevectors into an $\mu$-BLS spectrum. This additional mixing may affect marginally the amplitudes of the different spin wave contributions to an $\mu$-BLS spectrum. However since a large mixing is already inherent to Eq.~\ref{BLSspectrumMasterEquation}, the further mixing related to the existence of skew scattering does not invalidate the interpretation guideline to be used to discuss the characteristic features of a spectrum. The main determinants of an $\mu$-BLS spectrum will still be the dispersion relation of its spin wave eigenexcitations in the wavevector interval collected by the numerical aperture, the populations of the spin waves, as well as the interplay between their thickness profiles and the refractive optical properties of the sample.

\section{\label{sec:conclusion} Summary and conclusion}
By combining experiments and modeling, we have studied the method of microfocused Brillouin Light Scattering. We have built upon previous works to formulate a simple but comprehensive model
to explicitly calculate microfocused BLS spectra. The model takes into account the full complexity of the magnetic response expressed in the reciprocal space. This includes the dispersion relations of all spin wave families present in the sample, as well as their exact spatial profile within the thickness of the magnetic film, and their population. The model is written for systems whose equilibrium magnetization lies in the plane of the sample. 
The model also accounts for the main optical properties of the sample, and for the specificity of the experimental set-up (numerical aperture of the microscope objective, spectral resolution of the interferometer). By isolating the role of optical absorption, of precession ellipticity, and of the dispersive or non-dispersive character of spin waves, our model provides guidelines that can be used to interpret experimental microfocused BLS spectra.

The limitations of the model are the following. Photon skew scattering is ignored. Multiple reflections of the light within the material are not accounted for. The light absorption by non-magnetic capping layers is disregarded. The quadratic magneto-optical Kerr effects are neglected. Despite these approximations, the model proved accurate when performing a precise comparison with experiments done on a thick CoFeB magnetic film in which the spin waves are at thermal equilibrium.

\begin{acknowledgments}
This work was supported by the French National Research Agency (ANR) as part of the “Investissements d’Avenir” and France 2030 programs. This includes the SPICY project of the Labex NanoSaclay: ANR-10-LABX-0035, the MAXSAW project ANR-20-CE24-0025 and the PEPR SPIN projects ANR 22 EXSP 0008 and ANR 22 EXSP 0004. \\ We thank Jamal Ben Youssef and Aurélie Solignac for supplying BiYIG and CoFeB/Ru/CoFeB samples on which we could also test our theory. We thank Joo-Von Kim, Titiksha Srivastava and Eloi Haltz for their helpful comments, and Xavier Lafosse for the measurements of the optical properties of the sample.

\end{acknowledgments}

\section*{Appendix A: Normalization of the eigenvectors}
$\vec{\tilde{m}}_{\Psi_k}(k_x, k_y, z)$ is an eigenvector of magnetization dynamics. Being an eigenvector, it is defined up to an arbitrary multiplicative number. However for a correct description of the BLS signal, each $\vec{\tilde{m}}_{\Psi_k}(k_x, k_y, z)$ should be \textit{normalized} to correspond to the dynamic magnetization of a \textit{single} magnon per some given volume. This appendix describes the normalization procedure when using the outputs of TetraX.

The first step is to define the Holstein-Primakov variables $\tilde b_{\Psi k}$ associated to the dynamic magnetization profile of $\Psi_k$. 
If we write $\ell$ the cell index in the $z$ direction, the magnetization variables ($\vec m_0$, $\vec{\tilde{m}}_{\Psi_k}(k_x, k_y, z)$) are now defined for each $\ell$ and we can express the complex spin wave amplitude $\tilde b_{\Psi k}$ within a column of cells as \cite{korber_symmetry_2021}:
\begin{equation} 
\tilde b_{\Psi k} =  \sum_{\ell}  \big( \vec m_{0 }\times \vec{\tilde{m}}_{\Psi_k} \big).\vec{\tilde{m}}_{\Psi_k}^*
\end{equation}

The normalization is then conducted by checking that the measure of the transverse spin wave amplitude expressed by $\tilde b_{\Psi k}$ yields a population of exactly \textit{one} unit of longitudinal magnetization per column of cells, by scaling the magnetization profiles $\vec{\tilde{m}}_{\Psi_k}(k_x, k_y, z)$  such that:
\begin{equation} \eta(\Psi_k)= \tilde b_{\Psi k}^* \tilde b_{\Psi k} = 1\end{equation}
To ease the comparison of mode profiles, the phase of $\tilde m_{\Psi k}$ is chosen to ensure $\tilde m_{\Psi_k,z}(z=0) \in \mathbb R^+$.

\section*{Appendix B: contribution of the skew scattering}
The quantitative accounting of skew scattering is beyond the scope of this paper, but its effect can be qualitatively evaluated. Brillouin scattering depends on the respective orientations of light polarization and dynamic magnetization. When working with crossed input and output polarisers --like practiced in magnetic studies-- the source of scattering is the finite off-diagonal dielectric polarization $\vec P$ induced by the light within the material, which at first order is $\vec P \propto \tilde \epsilon_{xy} \vec E \times \vec {\tilde{m}}_\textrm{dyn}$ or:
\begin{equation}
\vec P \propto  \{E_y {\tilde{m}}_\textrm{dyn}^z-E_z {\tilde{m}}_\textrm{dyn}^y, -E_x {\tilde{m}}_\textrm{dyn}^z, E_x {\tilde{m}}_\textrm{dyn}^y \}. \label{DielectricPolarization}
\end{equation}
Eq.~\ref{DielectricPolarization} recalls that when $\vec E \parallel \vec {\tilde{m}}_\textrm{dyn}^z$ or when $\vec E \parallel \vec {\tilde{m}}_\textrm{dyn}^\phi$, the corresponding induced dielectric polarization simply vanishes. There, the absence of dielectric polarization suppresses all magnetic scattering: both the back-scattering and the skew scattering. This is for instance the case in Eq.~\ref{tildeL} when the in-plane component of the dynamical of the magnetization is parallel to the electric field, which translates as $\phi_k=\phi_0$ for our configuration.

However an additional phenomenon can lead to the absence of scattering of the photons towards a given direction, even when $\vec P \neq 0$. The skew scattering can either be enabled by the departing away from the back-scattering direction, or can be disfavored by this departing. This arises because light emission in a direction parallel to $\vec P$ is \textit{impossible} (it is a blind spot of the antenna diagram of the Hertzian dipole $\vec P$) while it is \textit{possible} in any other direction; the emission lobe is axially symmetric about $\vec P$ and it is maximal in the plane perpendicular to $\vec P$.

The contribution of skew scattering thus depends on each experimental situation. 
The out-of-plane part of the dielectric dipole (i.e. $P_z= E_x {\tilde{m}}_\textrm{dyn}^y$) has an isotropic emission diagram in the $xy$ plane, such that the back-scattering direction is as equally probable as any other skew scattering direction, which are all collected by the objective lens. This part of the $\mu$-BLS signal is thus not altered by the existence of skew scattering. 

Thanks to the dominant\footnote{In our case we have $\vec E_\textrm{inc} \parallel \vec e_x$, such that for most rays of incoming light, the electric field is mainly along $x$. At large (grazing) incidence the electric field can also have small $y$ and $z$ components, that we shall consider as of secondary importance, hence neglected, for skew scattering.} orientation of the electric field (i.e. $E_x \gg E_y, E_z$), the other large contributions to the $\mu$-BLS signals originate from $P_y=-E_x {\tilde{m}}_\textrm{dyn}^z$, and generalize the scattering situations depicted either in Figs.~\ref{calBLSDEBV}(a)-\ref{shcmconvt}(a), or in Fig.~\ref{calBLSDEBV}(b)-\ref{shcmconvt}(b). \\
In the first case ($\vec k \parallel \vec E \parallel \vec e_x $, generalizing Fig. \ref{calBLSDEBV}(a)), the skew scattering channels reroute outgoing light rays out of the (x-directed) back-scattering geometry, which was the optimum direction of emission of the Hertzian dipole $P_y$. The skew scattering channels thus render the system less sensitive to the out-of-plane dynamic magnetization. \\In the second case ($\vec k \perp \vec E$ and $\vec E \parallel \vec e_x $, generalizing Fig. \ref{calBLSDEBV}(b)), the skew scattering channels render the system more sensitive to out-of-plane dynamic magnetization because they reroute outgoing light rays towards of the optimum direction of emission of the Hertzian dipole $P_y$. 

These two cases show that the existence of skew scattering (in addition to back-scattering) tends to even further mix the contributions of the different SW wavevectors into an $\mu$-BLS spectrum. This additional mixing may affect marginally the amplitudes of the different spin wave contributions to an $\mu$-BLS spectrum.

\newpage 
$$   ~~~   $$
\newpage 
$$   ~~~   $$

\bibliography{article}
\newpage
\section{Supplementary: Dispersion relations in the Kalinikos and Slavin model}
\begin{figure}
    \centering
    \includegraphics[scale=0.1,width =8cm]{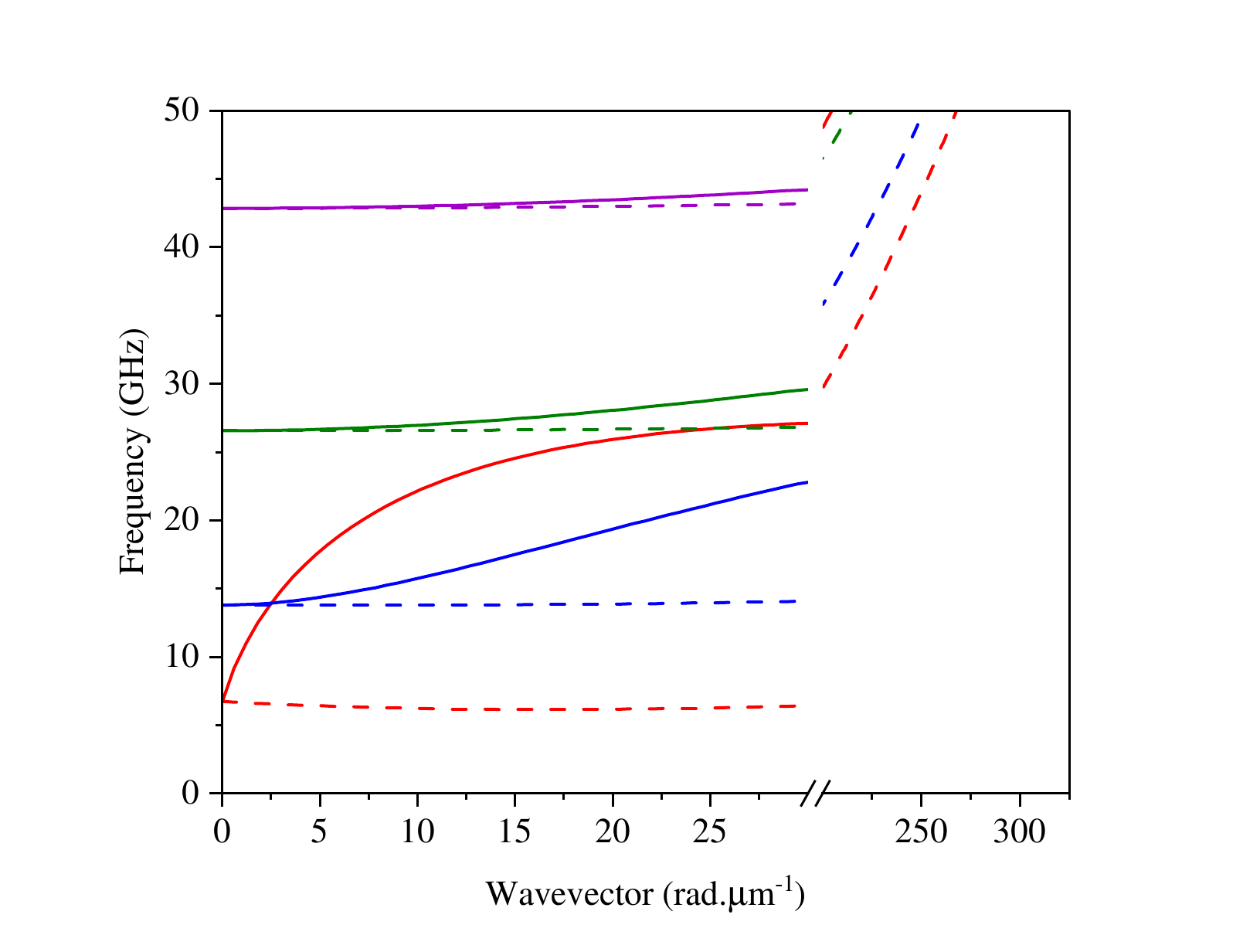} 
    \caption{Spin wave dispersion relations according to Kalinikos and Slavin model \cite{kalinikos_theory_1986} in the Damon-Eshbach geometry (bold lines, $\vec k \perp \vec m_0$) and the backward volume geometry (dashed lines, $\vec k \parallel \vec m_0$) for a field of 30 mT.}
\label{dispKS}
\end{figure}

The figure~\ref{dispKS} plots the dispersion relations of the backward-volume magnetostatic spin wave (BVSW), magnetostatic surface spinwaves (MSSW) and the three PSSW modes that exist below 50 GHz for wavevectors $||\vec k||$ up to $300~\textrm{rad}/\mu \textrm{m}$, as calculated using the Kalinikos and Slavin simplified formulas \cite{kalinikos_theory_1986}. A comparison with their equivalent ones calculated in the main document with TetraX \cite{korber_finite-element_2021} (Fig.~3(a)) evidences that the Kalinikos and Slavin expressions predict dispersion relations that are substantially more dispersive when $\vec k \perp \vec m_0$. This pronounced dispersive character significantly attenuates the Van Hove singularities in the SW density of states, as shown in the Fig.~3(b) of the main document. Moreover, the non-accounting of the mode hybridization within the analytical dispersion relations is such that the $p=0$ and $p=1$ bands simply cross instead of undergoing anticrossing. Since the spin wave density of states largely determines the shape of the microfocused BLS spectra, the analytical expressions of ref. \cite{kalinikos_theory_1986} can unfortunately not be used for the quantitative evaluation of the $\mu$-BLS spectra, at least for our chosen material (CoFeB, 50 nm thickness).

\section{Supplementary: Possible shapes of the microfocused BLS spectra at constant film thickness over exchange length ratio}
The goal of the section is to examine if one of the possible sources of disagreement between the model and the experimental spectrum could come from the values of the magnetic properties listed in Table III (magnetization, damping and exchange stiffness). Since the frequencies of the peaks of the model and the experiments match for all applied fields, the Zeeman interaction is correctly accounted for and the value of the magnetization must be correct. As the linewidths of the high frequency (narrow) peaks are determined by the spectral resolution of the setup, the value of the Gilbert damping does not affect the spectra and cannot be confirmed. The remaining parameter is the exchange stiffness $A_\textrm{ex}$. To identify its effect on the shape of the spectra without shifting the frequencies of the signal peaks, we ensure that the frequencies of the PSSW modes at $\vec k = \vec0$ stay unchanged. These frequencies follow the approximate expression \cite{hillebrands_spin_2002}:
\begin{equation}
     \omega_{k=0}=  \sqrt{(\omega_{H}+\omega_{M} \lambda_\textrm{ex}^2 q_v^2)(\omega_{H}+\omega_{M}+\omega_{M}\lambda_\textrm{ex}^2 q_v^2)}
     \label{fkittelpssw}
 \end{equation}
where $\omega_M=\gamma \mu_0 M_s $, $\omega_H=\gamma \mu_0 H$, $\lambda_\textrm{ex}=\sqrt{\frac{2A_{\textrm{ex}}}{\mu_0M_{s}^2}}$ and $q_{v}=v\frac{\pi}{t}$ with $v$ an integer.
We fix the frequencies $\omega_{k=0}$ by varying $\{A_\textrm{ex},t\}$ while keeping the ratio $\frac{A_{\textrm{ex}}}{t^2}$ constant. This modifies the dispersion relations, especially when $\vec k \perp \vec m_0$.
\subsection{Effect on the dispersion relations}
 
\begin{figure}
    \centering
    \includegraphics[scale=0.1,width =8cm]{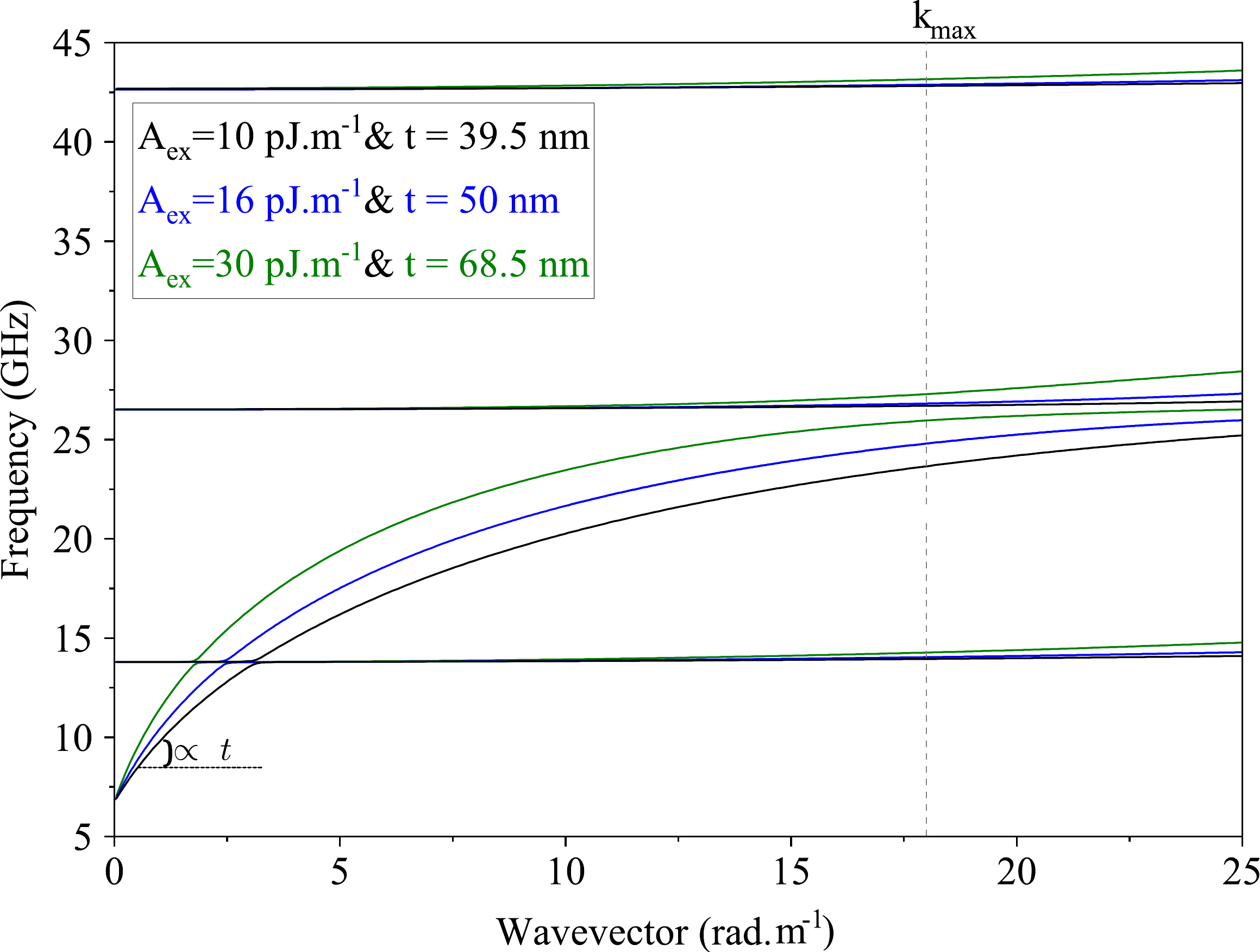}
    \caption{ Spin wave dispersion relations according to TetraX in the Damon-Eshbach geometry ( $\vec k \perp \vec m_0$)}
    \label{RDAex}

\end{figure}

Fig.~\ref {RDAex} plots the dispersion relation for $\vec k \perp \vec m_0$ for three $\{A_\textrm{ex},t\}$ pairs. 
Only the very dispersive Magnetostatic Surface Spin Wave (MSSW, "Damon-Eshbach" mode) undergoes substantial changes. At $k=0$, this mode has a group velocity \cite{yu_high_2012} of $\frac{\partial \omega}{\partial k_y}=  \gamma_0^2 M_s^2 t / (4 \omega_\textrm{FMR})$ which scales with the thickness.  As a result, the highest pair $\{30~\text{pJ}.\text{m}^{-1}$,~68.5 nm\} leads to higher frequencies than the other pairs. Consequently, the hybridization of the dispersive MSSW mode and the PSSW1 $v$ = 1 mode at 13.8 GHz occurs at a lower wavevector. \\ 
Besides, a second anticrossing is expected around 26 GHz, when the dispersive MSSW mode meets the PSSW2 mode. For our material parameters (Table III of the main document), this anticrossing should happen at a wavevector $k^y_\textrm{crossing} > k_\textrm{max}$, i.e. not accessible in the experiments. However choosing the highest value of exchange would potentially reduce $k^y_\textrm{crossing}$ and make it sensed, with consequences on the shape of the $p=2$ peak in the $\mu$-BLS spectrum. 

\begin{figure}
    \centering
    \includegraphics[width =8.5 cm]{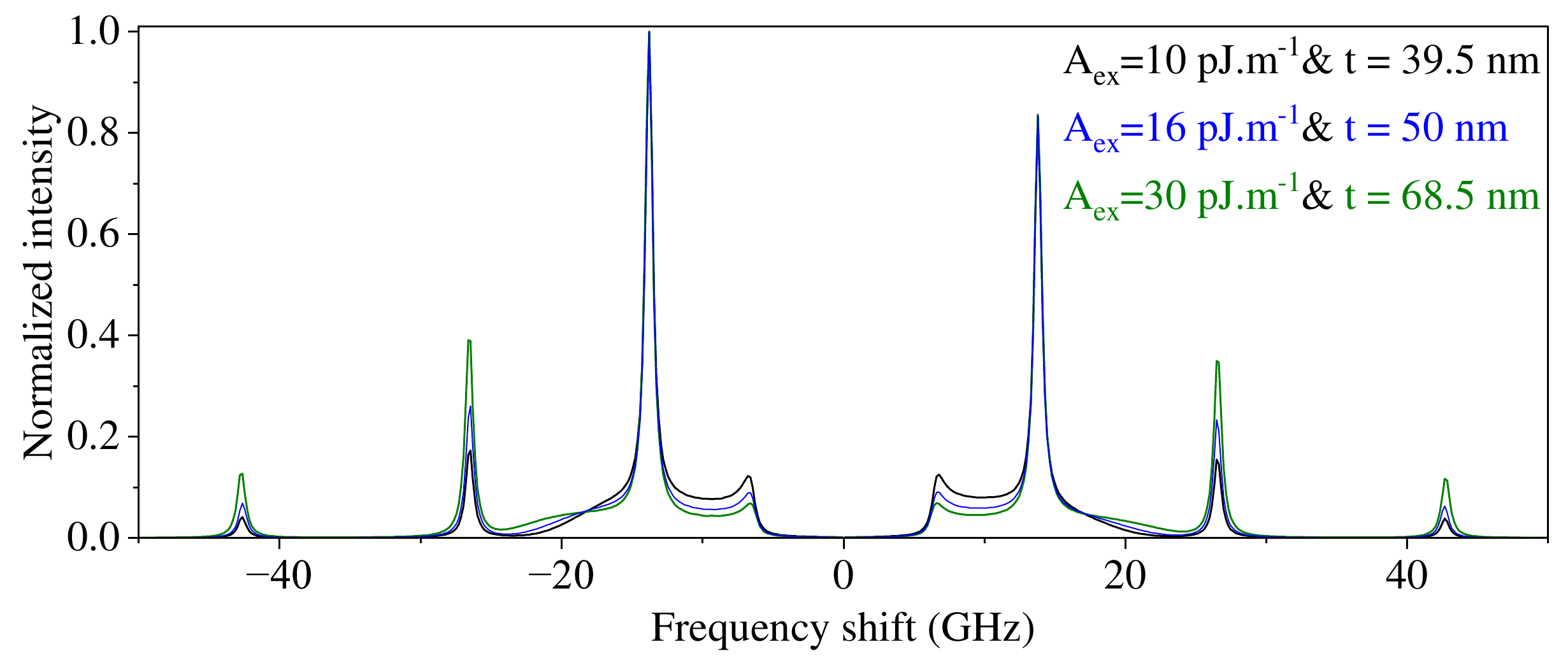} 
    \caption{Modeling $\mu$-BLS spectra for selected $A_\textrm{ex}$ and $t$. The spectra are normalized to the Stokes $p$=1 peak of the panel.  
    }
\label{fig_variationAex}
\end{figure}

The figure~\ref{fig_variationAex} presents the calculation of the $\mu$-BLS spectra for these three \{$A_{\text{ex}}$,t\} pairs.

\subsection{Effect on the lowest frequency peak}
Compared to the other pairs, the pair $\{10~\text{pJ}.\text{m}^{-1}$,~39.5 nm\} yields a higher signal at $\omega \approx \omega_\textrm{FMR}$ and slightly above that frequency; it also induces a faster roll-off at the frequency high end of this peak (circa 20 GHz). This can be understood from the dispersion relations (Fig.~\ref{RDAex}) and from Eq.~19 of the main document, which expresses the frequency $\frac{\omega}{2 \pi}\big(\textrm{MSSW},\,\phi_k=\frac \pi 2,\,k_\mathrm{max)}$ at which the $p=0$ (hill-like) peak rolls off. Increasing the thickness simply increases this frequency (see Fig.~\ref{RDAex}). Besides, the area of the $p=0$ peak is set by counting the spin wave states in the lowest band, with weakly varying weighting factors (see the main document). Since the total number of spin wave states in each band is thickness-independent, the spreading of the MSSW mode over a wider range of frequencies comes with a decrease in the amplitude of the peak maximum. 

\subsection{Effect on the high order peaks}
Increasing the thickness $t$ increases the amplitude of the $p \geq 2$ peaks in the $\mu$-BLS spectrum [Fig.~\ref{fig_variationAex}]. According to Eq.~10 in the main text, the amplitude of the peaks is set by the interplay between two physical phenomena. The first is the attenuation of the electromagnetic wave through the thickness. The second is the overlap between the electromagnetic wave (whose depth penetration is set by $k_\textrm{z, optic}$ =$\frac{4\pi \textrm{Re}(\tilde n)}{\lambda}$) and the spin wave thickness profile, which is essentially $k_\textrm{z,mag}$=$\frac{v\pi}{t}$ when far from the mode crossings, where $v$ is the number of nodes. We can express the overlap of these two waves by: 
\begin{equation}
\begin{split}
I'(z)& =e^{-\frac{4\pi \textrm{Im}(\tilde n) z}{\lambda}}\frac{1}{2}\Bigg[\cos \Bigg(k_{z,\textrm{optic}}z-k_\textrm{z,mag}z\Bigg)+\\
&\cos \Bigg(k_{z,\textrm{optic}}z+k_\textrm{z,mag}z\Bigg)\Bigg].
\end{split}
    \label{coscos}  
\end{equation}

\begin{figure}
    \centering
    \includegraphics[width =8cm]{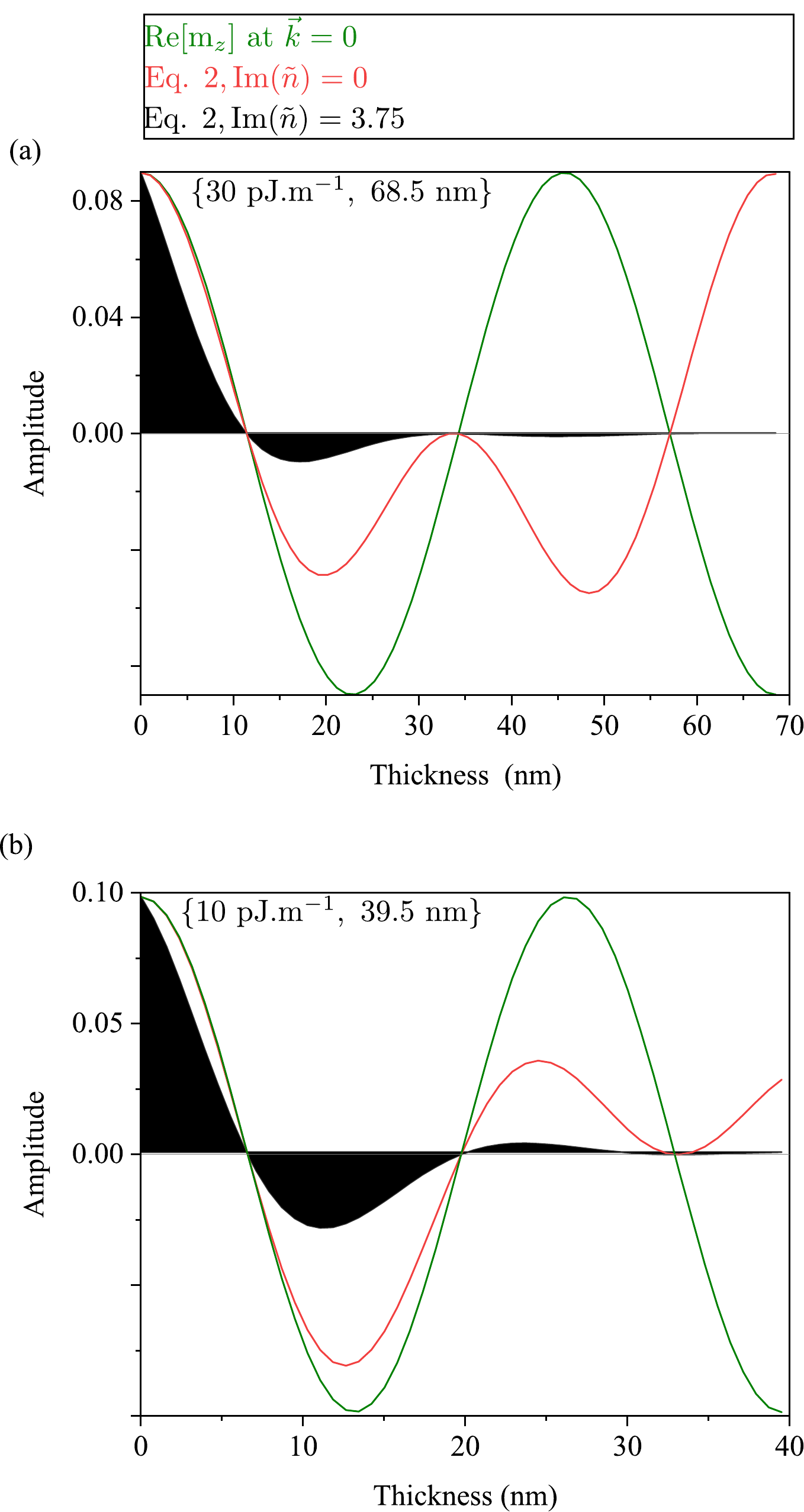}
    \caption{Comparison of the mode profile $\textrm{Re}\big(\tilde m_{z} (z) \big)$ of $p=3$  at $\vec k=\vec 0$ (green), with the results of Eq. \ref{coscos} with $\textrm{Im}(\tilde n)$=0 (red) and in black the results of Eq. \ref{coscos} with $\textrm{Im}(\tilde n)=3.75$ (black) for selected $\textrm{A}_{\textrm{ex}}$ and $t$. The magnetic parameters are $\mu_0 \text{M}_\text{s}$ = 1.77 T and (a)$\{30~\text{pJ}.\text{m}^{-1}$,~68.5 nm\} and (b) $\{10~\text{pJ}.\text{m}^{-1}$,~39.5 nm\}  }
    \label{modeprofilevselectromag}

\end{figure}

For a qualitative understanding, we have plotted in Fig.~\ref{modeprofilevselectromag} the mode profile of the third PSSW $v$ = 3 mode at $\vec k = 0$ and the results of Eq.~\ref{coscos} with and without light absorption. When $t=68.5$ nm [Fig.~\ref{modeprofilevselectromag}(a)], the quantity $I'(z)$ has a high amplitude positive crest followed by a much smaller trough, such that the integral of $I'(z)$ leads to a high $p=3$ BLS signal. For $t = 39.5$ nm [Fig.~\ref{modeprofilevselectromag} (b)], $I'(z)$ exhibits a crest followed by a trough having approximately the same amplitudes; the integral of $I'(z)$ thus reduces by partial cancellation. Equivalently, we can say that for a medium absorbing light, increasing the film thickness makes the PSSW modes look more uniform in the optically-sensed thickness, which increases the amplitude of their signature in $\mu$-BLS spectra. This explains why the amplitude of the $p \geq 2$ peaks in the $\mu$-BLS spectrum increases with the film thickness.

\section{Supplementary: Description of the notebook}
In this section, we describe how to use the notebook to calculate the $\mu$-BLS spectrum.
Once you have your magnetic input data, which are the frequency, the linewidth, and the mode profile for each in-plane wavevector \{$k_x,k_y$\},  the first step is to normalize the mode amplitudes. This is done by using the script named "normalization" which generates the normalization constant in a text file.  The second script "model-BLS" is the script allowing to generate the $\mu$-BLS spectra. 
These two Jupyter notebooks are written in Python. However, they are easy to use and do not require any prior knowledge of python programming. 
The example provided applies the script to a system of four modes families. However, if your system contains a different number of mode families, you simply need to adjust the corresponding part in the "Data input" section. 

The scripts are available at the following link:
https://github.com/nessrinebenaziz/model-BLS.git

\end{document}

%% file: table_input_model.tex
\begin{table}[]
    \centering
    \caption{Inputs required for the calculation of microfocused BLS spectra.}
\begin{tabular}{|m{4.5cm}|m{3.8cm}|}
  \hline
  Definition & Notation \\
  \hline
  \textbf{Magnetic inputs} & \\
  \hline
  Equilibrium magnetization & $\vec m_0(r)$ \\
  \hline
  Spin wave eigenmode & $\Psi_k$ \\
  \hline
  Spin wave wavevector & $\vec k$ \\
  \hline
  Spin wave frequency & $\omega_{\Psi_k} > 0$ \\
  \hline
  Spin wave population & $\eta_{\Psi_k}$ \\
  \hline
  Spin wave linewidth (FWHM) & $\Delta\omega_{\Psi_k}$ \\
  \hline
  Dynamic magnetisation of $\Psi_k$ & $\mathrm{Re}\big(\vec{\tilde{m}}_{\Psi_k} (r)e^{-i \omega_{\Psi_k} t}\big)$ \\
  \hline
  Incidence angle relevant for $\Psi_k$ & $\theta_k>0$ \\
  \hline
  Angle between the x-axis and SW wavevector & $\phi_k \in [0, 2\pi]$ \\
  \hline
  \textbf{Optical inputs} & \\
  \hline
  Diagonal term of the relative permittivity of the material & $\tilde \epsilon$ \\
  \hline
  Off diagonal term of the relative permittivity of the material & $\tilde \epsilon_{xy}$ \\
  \hline
\end{tabular}
\label{tab1}
\end{table}